\documentclass[aps,prl,twocolumn,reprint,floatfix, superscriptaddress ]{revtex4-2}
\usepackage[pdftex, colorlinks=true, linkcolor=blue, citecolor=blue, urlcolor=blue]{hyperref}

\usepackage{graphicx}% Include figure files
\usepackage{dcolumn}% Align table columns on decimal point
\usepackage{bm}% bold math
\usepackage{amssymb} % fancy caligraphic \Im
\usepackage{amsmath} % Matemáticas
\usepackage{amssymb} % Matemáticas
\usepackage{mathrsfs}
\usepackage{physics}
\usepackage{xcolor}
\usepackage{soul} % to cross text using \st{} command
\setstcolor{red}

\begin{document}

\preprint{APS/123-QED}

\title{Near-Field Gain and Far-Field Control via a Plasmonic Time Crystal Slab}% Force line breaks with \par

\author{Jaime E. Sustaeta-Osuna}
\affiliation{Departamento de F\'{i}sica Te\'orica de la Materia Condensada, Universidad Aut\'onoma de Madrid, E-28049 Madrid, Spain}
\affiliation{Condensed Matter Physics Center (IFIMAC), Universidad Aut\'onoma de Madrid, E-28049 Madrid, Spain}
\author{Thomas F. Allard}
\affiliation{Departamento de F\'{i}sica Te\'orica de la Materia Condensada, Universidad Aut\'onoma de Madrid, E-28049 Madrid, Spain}
\affiliation{Condensed Matter Physics Center (IFIMAC), Universidad Aut\'onoma de Madrid, E-28049 Madrid, Spain}
\author{Francisco J. García-Vidal}
\affiliation{Departamento de F\'{i}sica Te\'orica de la Materia Condensada, Universidad Aut\'onoma de Madrid, E-28049 Madrid, Spain}
\affiliation{Condensed Matter Physics Center (IFIMAC), Universidad Aut\'onoma de Madrid, E-28049 Madrid, Spain}
\author{Paloma A. Huidobro}
\affiliation{Departamento de F\'{i}sica Te\'orica de la Materia Condensada, Universidad Aut\'onoma de Madrid, E-28049 Madrid, Spain}
\affiliation{Condensed Matter Physics Center (IFIMAC), Universidad Aut\'onoma de Madrid, E-28049 Madrid, Spain}

\date{\today}% It is always \today, today,
             %  but any date may be explicitly specified

\begin{abstract}

Light-matter interactions can be substantially altered in the presence of time-varying media. We study the interaction between a harmonic electric dipole and a plasmonic time crystal slab. Temporal modulation of the plasma frequency enables near-field gain, allowing the dipole to absorb rather than emit energy, suppressing non-radiative losses. At the parametric resonance condition, the slab radiates strongly to the far-field, producing 100$\%$ oscillations in the radiated power at distances up to $10^3$ times the \textcolor{black}{epsilon-near-zero} wavelength. These findings reveal a new mechanism for controlling light–matter interaction in time-varying plasmonic systems.

\end{abstract}

%\keywords{Suggested keywords}%Use showkeys class option if keyword
                              %display desired
\maketitle
\textit{Introduction---} Periodic temporal modulations of a material optical properties lift up the constraints of passivity and energy conservation, leading to the emergence of photonic time crystals (PTCs) \cite{galiffi2022photonics}. These systems exhibit novel band structures with momentum gaps that host complex frequency modes \cite{reyes2015observation, galiffi2022photonics,boltasseva2024photonic,khurgin2024photonic}, and have been shown to amplify both classical electromagnetic waves \cite{galiffi2020wood,lyubarov2022amplified,gaxiola2023growing,wang2023metasurface,horsley2023eigenpulses} and the quantum vacuum \cite{mendoncca2000quantum, sloan2021casimir,sloan2022controlling,ganfornina2024quantum,sustaeta2025quantum}.
\par
Additionally, PTCs can also fundamentally modify light-matter interactions, such as the emission properties of classical dipoles and quantum emitters, as shown in Refs.  \cite{lyubarov2022amplified,dikopoltsev2022light,li2023stationary,park2024spontaneous,garg2025inverse,bae2025cavity,allard2025broadband}. However, most of these works have neglected material dispersion, and all have assumed a spatially infinite medium. Crucially, any realistic implementation of a PTC will be both dispersive and confined to a finite region of space. Thus, including both of these complexities and understanding their roles is necessary to fully understand light-matter interactions in time-periodic media.
\par
On the one hand, spatial confinement of PTCs to specific geometries, such as slabs \cite{holberg1966parametric,zurita2010resonances,reyes2012band,martinez2018parametric,galiffi2020wood,shirokova2023surface, globosits2024pseudounitary,valero2025resonant} or spheres \cite{ptitcyn2023floquet,verde2025optical}, shows that the lack of frequency conservation enables the different resonant modes of the structure to interact and hybridize, with the amplification phenomena displaying features that depend on the spatial symmetries of the structure (planar, spherical, etc.). On the other hand, material dispersion allows to engineer the size of the momentum gaps, which can be much larger in the dispersive case than in the non-dispersive one \cite{wang2025expanding,wang2025generating}. Moreover, periodically modulated plasmonic media 
(that is, \textit{plasmonic} time crystals), also support longitudinal modes. These can be amplified when the modulation frequency is resonant with the epsilon-near-zero (ENZ) frequency,
where $\Re\{\varepsilon(\omega_\text{ENZ})\}\sim 0$, giving rise to even larger momentum gaps than in the transverse case \cite{feinberg2025plasmonic}. Therefore, accounting for spatial confinement and material dispersion in a PTC is not only of practical importance to model real-world scenarios, but it also allows for a higher tunability and enhancement of the amplification phenomena supported by these time-periodic systems.
\par
In this Letter, we revisit the canonical setup for the study of nanoscale light-matter interactions, that is, dipole emission close to a nanostructure, but in the time-varying case. We analyze the emission properties of a harmonic electric dipole placed at a height $h$ above a plasmonic time crystal slab. We unveil new mechanisms for tailoring light-matter interactions both in the near-field  ($h$ smaller than the ENZ wavelength, $\lambda_\text{ENZ}$) and the far-field ($h \gg \lambda_\text{ENZ}$). We show that the time-modulation introduces gain into the system, allowing the dipole to absorb energy in the near-field. Therefore, the emitter ceases to be a source of electromagnetic radiation, instead becoming a sink. We also report very large (100$\%$ of the free-space value) and spatially long-lived oscillations of the radiated power in the far-field. \textcolor{black}{These oscillations arise from a sharp leaky resonance that is related to the Brewster condition and its negative frequency replica.}
\par
% made out of indium tin oxide (ITO) \cite{reshef2019nonlinear}. The choice of ITO is motivated by its prominent role as an experimental platform to realize time-varying phenomena \cite{alam2016large, vezzoli2018optical,zhou2020broadband,bohn2021all,bohn2021spatiotemporal,tirole2022saturable,tirole2023double,tirole2024second,galiffi2024optical,harwood2025space}
%

%ITO is a heavily doped semiconductor, whose optical response is dominated by the conduction band electrons and properly described with a Drude model \cite{reshef2019nonlinear,tirole2022saturable,tirole2023double,galiffi2024optical,harwood2025space,thouin2025field}

\begin{figure}
    \centering
    \includegraphics[width=1\linewidth]{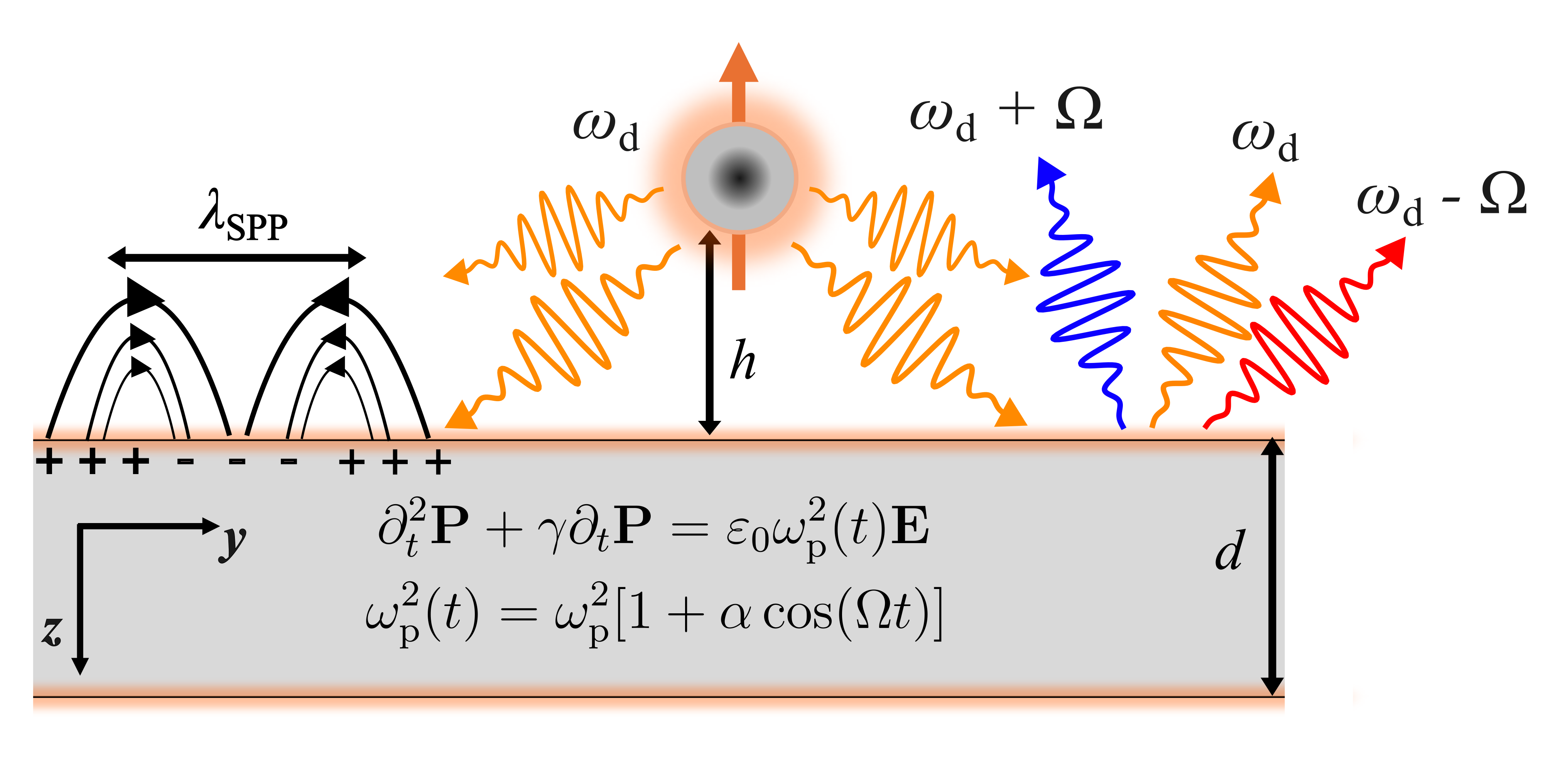}
    \caption{Diagram of the system under study: a harmonic electric dipole oscillating at frequency $\omega_{\text{d}}$, located at a height $h$ above a plasmonic time crystal slab. The slab supports surface plasmon polaritons. Additionally, owing to the lack of frequency conservation, waves can be reflected at different harmonics $\omega_{\text{d}} + n\Omega$, with $n$ an integer.}
    \label{fig:fig1}
\end{figure}

\textit{Physical model---}Our system is a periodically time-modulated \textcolor{black}{plasmonic} slab, as seen in Fig.~\ref{fig:fig1}. The time-modulation enters through the plasma frequency \cite{galiffi2024optical,harwood2025space,horsley2023eigenpulses}, which we assume is varied periodically in time as
\vspace{-0.0em}
\begin{equation}
    \omega^2_\text{p}(t) = \omega^2_\text{p}[1 + \alpha\cos(\Omega t)].
    \label{wp}
\end{equation}

In the above, $\Omega$ is the modulation frequency and $\alpha$, the modulation strength, while $\omega_\text{p}$ is the static (i.e., non-modulated) plasma frequency. The conduction band electrons thus obey the following parametric-Drude model equation
\vspace{-0.0cm}
\begin{equation}
    \partial^2_{t}\textbf{P} + \gamma\partial_t\textbf{P} = \varepsilon_0\omega_\text{p}^2(t)\textbf{E},
    \label{drude}
\end{equation}

\noindent where $\textbf{P}$ is the polarization density of the conduction electrons, $\gamma$ is their damping rate, $\textbf{E}$ is the externally applied electric field and $\varepsilon_0$, the permittivity of free space. In the above, we have omitted the time-dependence in $\textbf{P}$ and $\textbf{E}$, but have made explicit the one in $\omega_\text{p}(t)$ to emphasize that the system is being driven in time.

% We briefly mention in here that there is not a consensus as to how exactly the time-modulation should enter into the driven-Drude model equation, with some researches arguing that the damping rate too should be periodically varied in time (Simon's recent PRA, Diego Solís's paper). However, our choice of only modulating the plasma frequency, with the damping rate being constant, has a widespread use (Carsten papers, Simon's PRL with Emanuelle) and has been shown to match experimental results (CPA paper de Anthony, superluminal paper de Anthony), and therefore, our modeling of the system is perfectly adequate.
\par
The temporal periodicity of the plasma frequency allows us to use Floquet's theorem. \textcolor{black}{Therefore,} we make the following ansatzs: $\text{P}_{j}(t) = \sum_n\text{P}_{j, n}e^{-i\omega_nt}$ and $\text{E}_{j}(t) = \sum_n\text{E}_{j, n}e^{-i\omega_nt}$, where $j = x, \text{ }y\text{, }z$ labels the different Cartesian components and $\omega_n = \omega + n\Omega$. Here, $\omega$ is a real frequency and $n\in\mathbb{Z}$. Under these Floquet ansatzs, we can explicitly solve the parametric-Drude differential equation, with $\text{P}_{j, n} = \varepsilon_0\sum_{n'}\chi_{n,n'}\text{E}_{j,n'}$, where we introduce the following Floquet susceptibility matrix
\begin{equation}
    \begin{split}
        \chi_{n,n'} & = \Big(\varepsilon_\infty - 1 - \frac{\omega_\text{p}^2}{\omega_n^2 + i\gamma\omega_n}\Big)\delta_{n,n'} - \\&
        \frac{\alpha}{2}\frac{\omega_\text{p}^2}{\omega_n^2 + i\gamma\omega_n}\big(\delta_{n+1,n'} + \delta_{n-1,n'}\big).
    \end{split}
    \label{floquet_chi}
\end{equation}
The first line in Eq. \eqref{floquet_chi} corresponds to the static response of the system, while the second line contains the effects of the temporal modulation. Additionally, the background permittivity $\varepsilon_\infty$ (not captured by the driven-Drude equation, but still demanded by the Kramers-Kronig relations \cite{jackson2021classical}) has been included. The material response is thus dispersive, i.e., frequency dependent, and time-modulated, since the frequency domain susceptibility is not a scalar, but a matrix. Hence, the time-modulated linear response allows different harmonics to interact. %A crucial aspect of the Floquet susceptibility matrix is that it also couples positive and negative frequencies; the importance of this shall be made clear when we discuss the dipole radiation.
\par
%Next, we solve Maxwell equations $\curl\textbf{E} = -\partial_t\textbf{B}$ and $\curl\textbf{B} = (1/c_0^2)\partial_t\textbf{E} + \textbf{J}$ inside and outside the ITO slab, and apply the continuity of the tangential components of the electric and magnetic fields $\textbf{E}$ and $\textbf{B}$. Material response enters Maxwell equations inside the slab through the current $\textbf{J}=\partial_t\textbf{P}$, where $\textbf{P}$ is the polarization induced by the externally applied electric field, as defined above. 
Next, we solve Maxwell equations $\curl\textbf{E} = -\partial_t\textbf{B}$ and $\curl\textbf{H} = \partial_t\textbf{D}$, making use of the constitutive relations $\text{B}_{j,n} = \mu_0\text{H}_{j,n}$ and $\text{D}_{j,n} = \varepsilon_0\sum_{n'}\varepsilon_{n,n'}\text{E}_{j,n}$, where $\varepsilon_{n,n'} = \delta_{n,n'} + \chi_{n,n'}$ is the Floquet permittivity matrix. By imposing the boundary conditions at the slab interface for all harmonics [see the Supplemental Material (SM)\cite{supmat}], we calculate the reflected field for a given polarization $\sigma =$ TE, TM and lateral momentum $K_\parallel$, both of which are conserved quantities. Frequency, on the other hand, is only conserved modulo $\Omega$ and therefore, the reflected field will contain several harmonics $\omega_n = \omega + n\Omega$ \cite{ptitcyn2023floquet,globosits2024pseudounitary}. Therefore, the tangential component of the reflected field at the air-metal interface (for given $\sigma$ and $K_\parallel$) reads $\text{E}^{\text{(ref)}}_{\sigma}(t;K_\parallel) = \sum_n\text{E}^{\text{(ref)}}_{\sigma}(\omega_n,K_\parallel)e^{-i\omega_nt}$, where
\begin{equation}
    \text{E}^{\text{(ref)}}_{\sigma}(\omega_n,K_\parallel) = \sum_{n'}R_{\sigma}(\omega_n,\omega_{n'};K_\parallel)\text{E}^{\text{(in)}}_{\sigma}(\omega_{n'},K_\parallel),
\end{equation}
with $\text{E}^{\text{(in)}}_{\sigma}(\omega_{n'},K_\parallel)$ being the amplitudes of the tangential incident field. 
\par
We have introduced generalized Fresnel-Floquet reflection coefficients $R_{\sigma}(\omega_n,\omega_{n'};K_\parallel)$ that couple an incident frequency $\omega_{n'}$ with an outgoing one $\omega_n$, where $\omega_n - \omega_{n'} = (n - n')\Omega$. Since we shall deal with a harmonic electric dipole, which oscillates at a single frequency, we only discuss in here the frequency conserving coefficient $R_{\sigma}(\omega_0,\omega_{0};K_\parallel)$, and for $\sigma = \text{TM}$ in order to maximize dipole coupling. %(for a more detailed discussion, the interested reader can refer to the SM \cite{supmat}).
\par
\begin{figure}
\makebox[\columnwidth][c]{
    \centering
    \hspace{0 mm}
    \includegraphics[scale = 0.47]{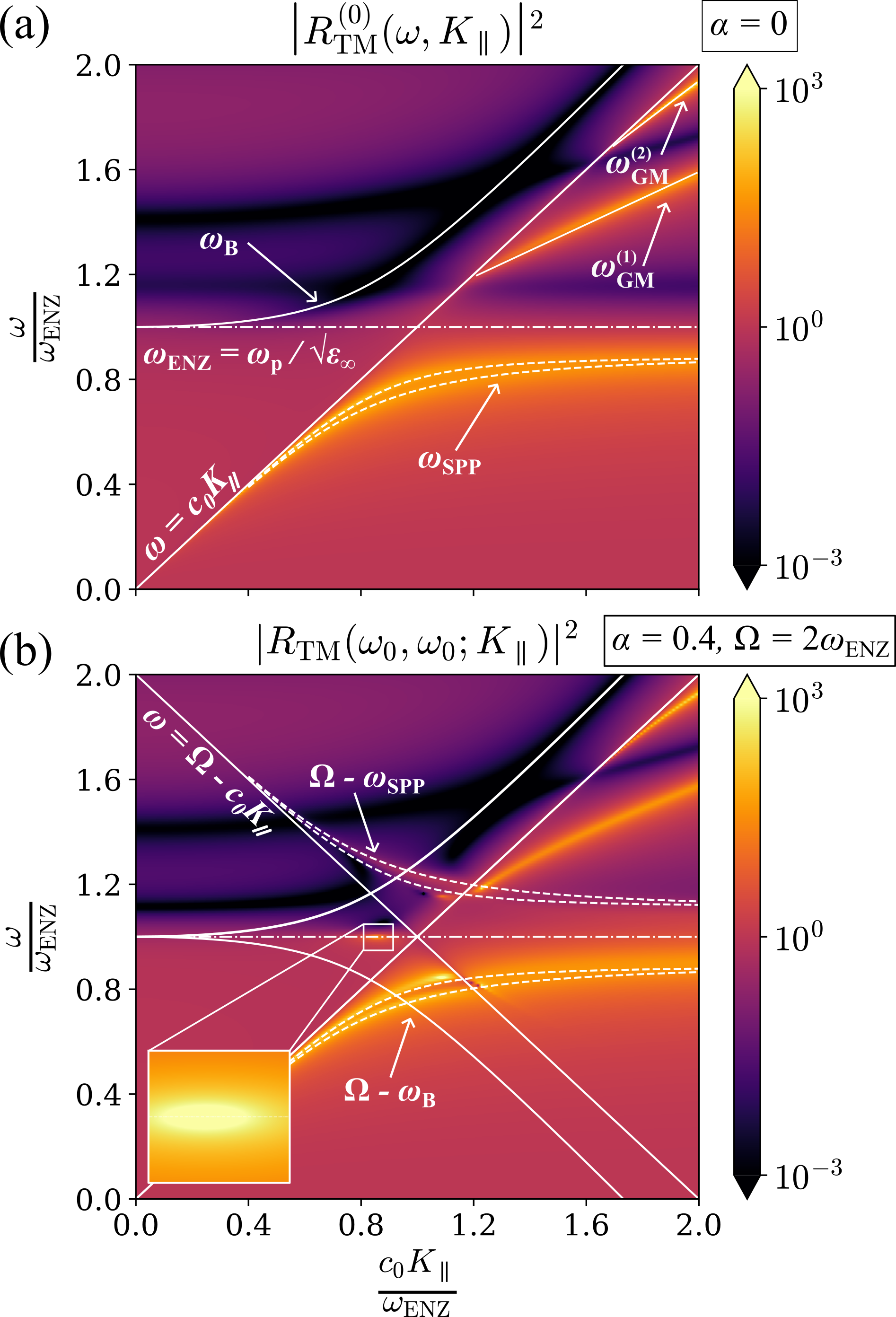} }
    \caption{Static (a) and time-modulated (b) TM reflectance vs. frequency and momentum. Temporal modulation generates negative-frequency replicas of the SPPs and the Brewster condition. Inset: zoom of the leaky resonance inside the lightcone at $\omega\sim0.5\Omega=\omega_\text{ENZ}$.}
    \label{fig:fig2} 
\end{figure}
\textit{Fresnel-Floquet reflection coefficient---} \textcolor{black}{In order to facilitate a quantitative discussion, we assume the slab to be made out of indium-tin-oxide (ITO)\cite{reshef2019nonlinear}. The optical response of ITO is dominated by the conduction band electrons and therefore, is properly described with a Drude model \cite{reshef2019nonlinear,thouin2025field}. Additionally, ITO serves as a standard experimental platform to realize ultrafast and ultrastrong temporal modulations of the electric permittivity \cite{alam2016large,bohn2021all,tirole2022saturable,harwood2025space,galiffi2024optical}}. The values for the Drude-model parameters and the slab's thickness are all taken from experimental data, in particular, those found in Ref.~\cite{galiffi2024optical}, with plasma wavelength $\lambda_\text{p} = 2\pi c_0/\omega_\text{p} = 654$ nm, damping rate $\gamma=0.13\text{ (fs)}^{-1}$, background permittivity $\varepsilon_\infty=4$ and thickness $d = 310$ nm. In Fig.~\ref{fig:fig2}, we compare the static reflectance for TM-polarized waves, $|R^{(0)}_{\text{TM}}(\omega,K_\parallel)|^2$ [panel (a)], with its time-modulated counterpart, $|R_{\text{TM}}(\omega_0,\omega_0;K_\parallel)|^2$ [panel (b)]. The vertical axis for both panels is the frequency of the incident and reflected waves, $\omega$, normalized to the ENZ frequency $\omega_{\text{ENZ}} = \omega_\text{p}/\sqrt{\varepsilon_\infty}$, whereas the horizontal axis is the lateral momentum $K_\parallel$, normalized to $c_0 / \omega_{\text{ENZ}}$.
%At $\omega = \omega_\text{ENZ}$ we have $\Re\{\varepsilon\} \sim 0$. Hence, at the ENZ frequency the electric field is purely longitudinal and the system enters a quasistatic regime, in which the spatial and temporal degrees of freedom decouple \cite{liberal2016nonradiating,gong2022radiative}
\par
In Fig.~\ref{fig:fig2}(a), we highlight both the lightcone (solid white line) and the \textcolor{black}{ENZ} frequency (horizontal dashed-dotted white line), which separates the metallic ($\omega<\omega_{\text{ENZ}}$) and insulating regimes ($\omega>\omega_{\text{ENZ}}$). We also mark with dashed white lines the surface plasmon polaritons (SPPs) supported by the metallic slab, $\omega_{\text{SPP}}$ \cite{ritchie1957plasma,otto1968excitation,barnes2003surface,greffet2012introduction,maier2007plasmonics,novotny2012principles}; these modes are confined to the air-metal interface and propagate along it \cite{greffet2012introduction,picardi2017unidirectional}, and appear as poles of the reflection coefficient for TM-polarized waves \cite{novotny2012principles,kim2025complex}. %Importantly, coupling to them results in an exponential increase of the total power radiated by a dipole as the air-metal interface is approached.
\par
For $K_\parallel$ large enough, we reach the quasistatic regime, in which the two SPPs converge to the surface plasmon (SP) \cite{ritchie1957plasma}, whose frequency is $\omega_{\text{SP}} = \omega_\text{p} / \sqrt{1 + \varepsilon_\infty}$. The latter gives a flat ($K_\parallel-$independent) resonance which, in the static case, greatly enhances the quenching of power radiated by a dipole placed in the near field of the slab \cite{novotny2012principles}. Additionally, above the \textcolor{black}{ENZ} frequency we can see the first two guided modes (GMs) of the slab \cite{jackson2021classical}, $\omega^{(1)}_{\text{GM}}$ and $\omega^{(2)}_{\text{GM}}$, highlighted in solid white, as well as the Brewster condition $\omega_\text{B}$\textcolor{black}{, also marked in solid white, } at which the reflection coefficient \textcolor{black}{for TM-polarization} vanishes.
\par
In Fig.~\ref{fig:fig2}(b) we focus on the time-modulated case, with values of $\alpha=0.4$ and $\Omega = 2\omega_{\text{ENZ}}$. We highlight the upshifted negative frequency replicas of the SPPs, $\Omega - \omega_{\text{SPP}}$, which are seen to interact with $\omega_\text{B}$. %We see that $\Omega - \omega_{\text{SPP}}$ anti-crosses both the Brewster angle and $\omega^{(1)}_{\text{GM}}$, whereas $\Omega - \omega^{(1)}_\text{GM}$ anti-crosses $\omega_\text{SPP}$; these anti-crossings result in the opening of $K_\parallel-$gaps, as seen in the SM \cite{supmat}.
\textcolor{black}{We also mark the upshifted negative frequency replica of the Brewster condition, $\Omega - \omega_\text{B}$. Surprisingly, the temporal modulation makes both the reflectivity poles and zeroes of the static slab to interact, with the zeroes effectively behaving as modes of the system. This is indeed the case for $\omega_\text{B}$ and $\Omega - \omega_\text{B}$, which interact at $\omega = 0.5\Omega = \omega_\text{ENZ}$ and $K_\parallel = 0$. This interaction results in the formation of a sharp leaky resonance inside the lightcone at $\omega = \omega_\text{ENZ}$, which appears as a peak in the inset of Fig.~\ref{fig:fig2}(b). The resonance is first formed at $K_\parallel = 0$ and moves towards larger values of $K_\parallel$ as $\alpha$ increases, resembling the behavior of the edge of a $K_\parallel-$gap (a detailed account of this can be found in the SM \cite{supmat}).}
%at the bulk plasmon frequency, the ITO slab acts as a resonator, with all conduction band electrons oscillating coherently and in phase at $\omega_\text{ENZ}$. Therefore, by modulating at $\Omega = 2\omega_\text{ENZ}$, we are driving the resonator at its parametric amplifier condition.
%and for sufficiently large $\alpha$, this can lead to an amplifying regime in which the slab radiates electromagnetic energy to the far-field \cite{supmat}.

% makes the bulk plasmon radiate electromagnetic energy to the far-field, as discussed earlier. Hence, our choice of $\Omega = 2\omega_{\text{ENZ}}$ leads to interesting phenomena both in the near-field and the far-field.
\par
\begin{figure*}
\makebox[\columnwidth][c]{
    \centering
    \hspace{-0.5 cm}
    \includegraphics[scale =  0.43]{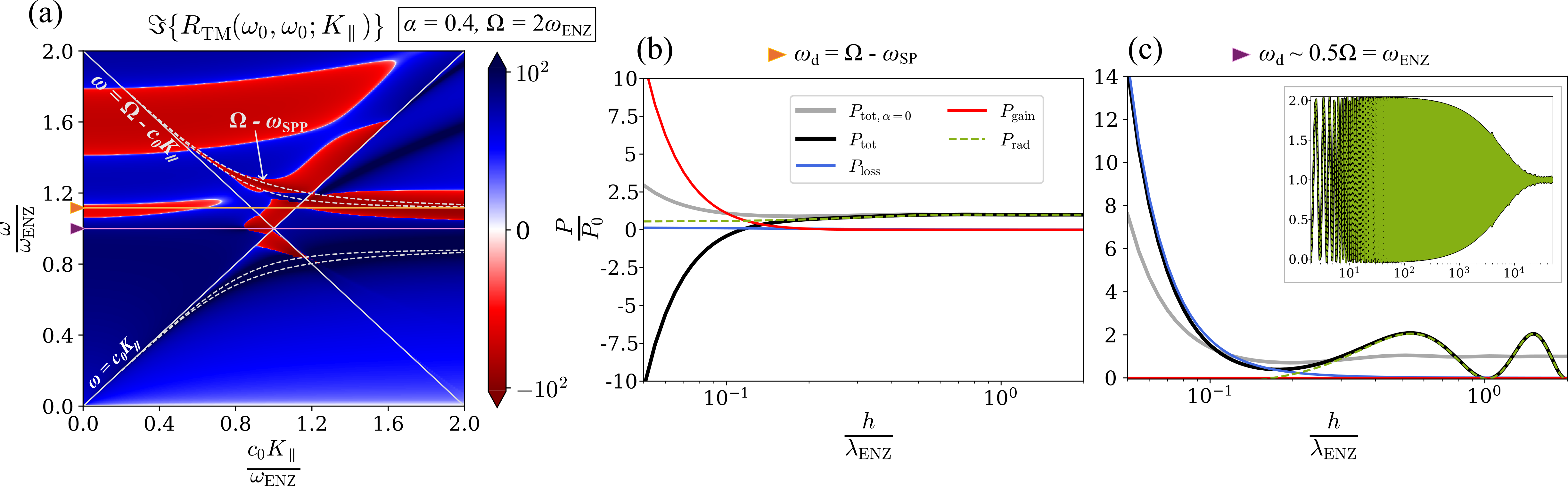} }
    \caption{(a) Imaginary part of the frequency-conserving TM reflection coefficient, versus frequency and momentum. Outside the lightcone, blue-colored regions correspond to near-field losses, while red-colored ones are associated to near-field gain. Adjacent panels display the total power radiated by the dipole as a function of its height above the surface, for $\omega_\text{d} = \Omega - \omega_\text{SP}$ (b) and $\omega_\text{d}\sim0.5\Omega$ (c). In (b), gain dominates the near-field and the total power is negative in the vicinity of the air-metal interface. Inset in (c) shows large (100$\%$) and long-lived (up to $10^3\lambda_\text{ENZ}$) far-field oscillations.}
    \label{fig:fig3} 
\end{figure*}
\textit{Dipole radiation---}We now place a harmonic electric dipole at a height $h$ above the plasmonic slab, as shown in Fig.~\ref{fig:fig1}. The dipole oscillates at a frequency $\omega_{\text{d}}$, and its dipole moment $\textbf{d}$ is assumed to be perpendicular to the air-metal interface to maximize the coupling to TM modes. The energy exchanged per unit time between the dipole and the electric field -- which includes both the dipole's own field and the one reflected from the surface -- is given by $P(t) = \int d^3x\textbf{J}_{\text{d}}\cdot\textbf{E}$, where $\textbf{J}_{\text{d}} = -i\omega_{\text{d}}\textbf{d}e^{-i\omega_{\text{d}}t}\delta(\textbf{x} - \textbf{x}_0)\text{ }+\text{ c.c.}$ is the current associated to the classical emitter. The field reflected from the plasmonic slab contains many different harmonics $\omega_{\text{d}} + n\Omega$, as seen in Fig.~\ref{fig:fig1}. However, as the dipole is monochromatic, for $t \gg1/\omega_{\text{d}}$, only the fundamental one $n = 0$ has a non-vanishing contribution to the total radiated power $P_\text{tot} = \underset{t\longrightarrow\infty}{\lim}\frac{1}{t-t_0}\int^t_{t_0} dt'P(t')$ \cite{supmat, novotny2012principles}, where $t_0\leq t$ is arbitrary and determines the initial time at which the dipole starts radiating. The total power can be separated into its radiative and non-radiative contributions, with $P_\text{tot} = P_\text{rad} + P_\text{non-rad}$. Particularizing the latter two for $\textbf{d}$ perpendicular to the surface we have

\begin{equation}
    \frac{P_\text{rad}}{P_0}  = 1\hspace{0.00cm}+\hspace{0.00cm} \frac{3c_0^3}{2\omega_{\text{d}}^3}\hspace{-0.05cm}\int_0^{\frac{\omega_{\text{d}}}{c_0}}\hspace{-0.1cm}\frac{dK_\parallel K_\parallel^3}{k_z}\mathfrak{R}\left\{R_{\text{TM}}(\omega_{\text{d}}, \omega_{\text{d}};K_\parallel)e^{-i2k_zh}\right\}
    \label{eq:prad}
\end{equation}

and

\begin{equation}
    \frac{P_\text{non-rad}}{P_0} = \frac{3c_0^3}{2\omega_{\text{d}}^3}\int_{\frac{\omega_{\text{d}}}{c_0}}^\infty \hspace{-0.1cm}\frac{dK_\parallel K_\parallel^3}{\kappa_z}\mathfrak{I}\left\{R_{\text{TM}}(\omega_{\text{d}}, \omega_{\text{d}};K_\parallel)\right\}e^{-2\kappa_zh}\hspace{-0.05cm}.
    \label{eq:pnon-rad}
\end{equation}

In Eqs.~\eqref{eq:prad} and \eqref{eq:pnon-rad}, $P_0 = \mu_0\abs{\textbf{d}}^2\omega_{\text{d}}^4/(3\pi c_0^3)$ is Larmor's formula, which gives the total radiated power in free-space and to which we normalize $P_\text{rad}$ and $P_\text{non-rad}$. On the other hand, $k_z = \sqrt{\left(\omega_{\text{d}}/c_0\right)^2 - K^2_\parallel}$ and $\kappa_z = \sqrt{K^2_\parallel -\left(\omega_{\text{d}}/c_0\right)^2}$ are the wave-numbers for plane and evanescent waves along the $z-$axis, which we take to be perpendicular to the air-metal interface (see Fig. \ref{fig:fig1}). Thus, $P_\text{rad}$ is seen to be completely determined by the modes inside the lightcone, while $P_\text{non-rad}$ is given by the ones outside of it, such as the GMs or the SPPs.
\par
A glance at the formula for $P_\text{non-rad}$ shows that the integrand's sign is completely determined by the imaginary part of the generalized reflection coefficient at the dipole's frequency, \hspace{-0.0cm}$\mathfrak{I}\{R_{\text{TM}}(\omega_{\text{d}}, \omega_{\text{d}};K_\parallel)\}$. Therefore, having $\mathfrak{I}\{R_{\text{TM}}(\omega_{\text{d}}, \omega_{\text{d}};K_\parallel)\}\geq0$ enhances the power emitted by the dipole in the near-field with respect to the free-space value, with non-radiative losses dominating over radiative ones. Crucially, a negative value of $\mathfrak{I}\{R_{\sigma}(\omega_{\text{d}},\omega_{\text{d}};K_\parallel)\}$ yields a negative contribution $P_{\text{non-rad}}$. The latter is to be understood as the dipole absorbing energy instead of radiating it, with the electromagnetic field doing work on the dipole \textcolor{black}{instead of the opposite}. Bearing this in mind, we can further separate $P_{\text{non-rad}}$ into a loss contribution $P_{\text{loss}}$, determined by the $\mathfrak{I}\{R_{\text{TM}}(\omega_{\text{d}}, \omega_{\text{d}};K_\parallel)\}\geq0$ regions, and a gain one, $P_{\text{gain}}$, given by the $\mathfrak{I}\{R_{\text{TM}}(\omega_{\text{d}}, \omega_{\text{d}};K_\parallel)\}<0$ regions. Then, the non-radiative power is $P_\text{non-rad} = P_\text{loss} - P_{\text{gain}}$, where $P_{\text{gain}}$ is defined to be positive \cite{ren2024classical,park2024spontaneous}. Finally, summing radiative, loss and gain contributions, the total power reads $P_{\text{tot}} = P_{\text{rad}} + P_{\text{loss}} - P_{\text{gain}}$.
%For negative frequencies, this inequality becomes $\mathfrak{I}\{R^{(0)}_{\sigma}(-|\omega|; K_\parallel)\} \leq 0$ for $K_\parallel \geq |\omega| / c_0$, since negative-frequency fields are just complex conjugates of positive-frequency ones. Consequently, the imaginary part of the reflection coefficient flips sign under frequency inversion: $\mathfrak{I}\{R^{(0)}_{\sigma}(-\omega; K_\parallel)\} = -\mathfrak{I}\{R^{(0)}_{\sigma}(\omega; K_\parallel)\}$.
Notice that, in passive media, and owing to causality, the reflection coefficient has a non-negative imaginary part outside the light cone, with $\mathfrak{I}\{R^{(0)}_{\sigma}(\omega; K_\parallel)\} \geq 0$ for $K_\parallel \geq \omega / c_0$. Therefore, $P_\text{gain} = 0$, and the non-radiative power is necessarily positive. In media with gain, this restriction is lifted, and it thus becomes possible to have $\mathfrak{I}\{R_{\sigma}(\omega_{\text{0}},\omega_{\text{0}};K_\parallel)\} < 0$ for $K_\parallel \geq \omega / c_0$ and consequently, $P_\text{gain} > 0$.
\par
In Fig.~\ref{fig:fig3}(a) we plot $\mathfrak{I}\{R_{\text{TM}}(\omega_{0},\omega_{0};K_\parallel)\}$ as a function of the frequency $\omega$ and lateral momentum $K_\parallel$, again for $\alpha=0.4$ and $\Omega = 2\omega_\text{ENZ}$; blue-colored regions have $\mathfrak{I}\{R_{\text{TM}}(\omega_{\text{0}},\omega_{\text{0}};K_\parallel)\}>0$, while red-colored ones have $\mathfrak{I}\{R_{\text{TM}}(\omega_{\text{0}},\omega_{\text{0}};K_\parallel)\}<0$. According to the preceding discussion, we can identify blue regions outside the lightcone with loss and red ones with gain.
\par
\textit{Near-Field \hspace{-0.0cm}Gain---}In Fig.~\ref{fig:fig3}(a), we can see that $\mathfrak{I}\{R_{\text{TM}}(\omega_{\text{0}},\omega_{\text{0}};K_\parallel)\}$ takes negative values along the $\Omega - \omega_{\text{SPP}}$'s dispersion curves, which appear as red-colored in the figure. The SP replica (the quasistatic limit of $\Omega - \omega_{\text{SPP}}$) is particularly interesting, since a dipole whose frequency is matched to $\Omega - \omega_\text{SP}$ will couple to a flat region of gain, absorbing energy instead of radiating it. More importantly, the closer the emitter is to the surface, the more it will absorb. In Fig.~\ref{fig:fig1}(b) we plot the power $P$ (normalized to its free-space value $P_0$) versus the dipole's height above the surface $h$ (normalized to the ENZ wavelength $\lambda_{\text{ENZ}} = \sqrt{\varepsilon_\infty}\lambda_p$), and for $\omega_\text{d} = \Omega - \omega_\text{SP}$. The solid grey line corresponds to total power in the static case, $P_{\text{tot, } \alpha=0}$, while the solid black one is the total power $P_{\text{tot}}$ for $\alpha=0.4$ and $\Omega = 2\omega_\text{ENZ}$. We also represent the different contributions to the total power: $P_\text{rad}$ in dashed green, $P_\text{loss}$ in solid blue and $P_\text{gain}$ in solid red. We see that $P_\text{loss}$ vanishes completely, while $P_\text{gain}$ indeed becomes increasingly large as $h$ becomes smaller. Consequently, $P_\text{tot}$ decreases as the air-metal interface is approached, eventually becoming negative in the near-field and reaching values of $P_\text{tot}<-10P_0$ for $h = 0.05\lambda_\text{ENZ}$. 
%In the static case, there are two contributions to the near-field radiated power. The first one dominates at larger distances and comes from the SPP pole, giving a slight enhancement of the total power \cite{greffet2012introduction,maier2007plasmonics,novotny2012principles}. The second one dominates at shorter distances, in the quasistatic regime, and comes from non-radiative losses, which further enhance the dipole emission. For $\omega_\text{d} = \omega_\text{SP}$, such an enhancement becomes dramatic owing to the dipole coupling to the surface plasmon \cite{greffet2012introduction,maier2007plasmonics,novotny2012principles}.
\par
%Interestingly, similar phenomena occur in the time-modulated case when $\omega_\text{d} = \Omega - \omega_\text{SP}$. 
Crucially, it is at the transition from the SPP-replica dominating regime to the quasistatic one where $P_\text{gain}$ begins to grow large and correspondingly, where $P_\text{tot}$ becomes negative, as shown in the SM \cite{supmat}. Therefore, the quenching of the radiated power at $\omega_\text{SP}$ is transformed by the time-modulation into non-radiative gain at $\Omega - \omega_\text{SP}$, with non-radiative losses becoming completely suppressed in the process.
%We see $P_\text{loss}$ vanishes completely, while $P_\text{gain}$ has an exponential increase followed by a quasistatic ($~1/h^3$) contribution. Correspondingly, the total power decreases exponentially as the surface is approached, eventually becoming negative and reaching values of $P_{\text{tot}} < -750P_{0}$. Therefore, for $\omega_{\text{d}} = \Omega - \omega_{SP}$, non-radiative losses are completely suppressed, with non-radiative gain completely dominating in the near-field.
\par
%We also comment further about the $\Omega\gtrsim2\omega_{\text{SP}}$ condition: owing to the exponential decay present in $P_\text{tot}$'s formula, we need $\Omega - \omega_\text{SP}$ to lie close to the lightcone or otherwise, its contribution will be much diminished or even killed of by the exponential. Having $\Omega\gtrsim2\omega_{\text{SP}}$ precisely solves this issue and the SP replica can therefore be efficiently probed by the dipole. %Additionally, having non-zero $\gamma$ means the SPPs are not sharp resonances and have instead a finite width; this turns out to have a positive effect, since it extends the gain condition beyond $\omega_{\text{d}} = \Omega - \omega_{\text{SP}}$ \cite{supmat}.
%\par
\textit{Far-Field Control---}In Fig.~\ref{fig:fig3}(c) we study the $\omega_{\text{d}} \sim 0.5\Omega = \omega_\text{ENZ}$ case. As expected from Fig.~\ref{fig:fig3}(a), the non-radiative gain is zero and the near-field is dominated by non-radiative losses, which are enhanced by a factor of two with respect to the static case. The far-field shows unexpected features, with $P_\text{tot}$ displaying large and spatially long-lived oscillations that are a 100$\%$ of the free-space value. These oscillations stem from the \textcolor{black}{sharp peak inside the lightcone at $\omega = \omega_\text{ENZ}$ [see inset of Fig.~\ref{fig:fig2}(b)], which is rooted in the interaction between $\omega_\text{B}$ and $\Omega - \omega_\text{B}$.} %This amplifying regime has features similar to those of a lasing phenomenon, and is described in detail in the SM \cite{supmat}.
\par
In this way, the emission of light by the dipole can be enhanced by a factor of two or be completely suppressed, depending on the value of $h$, \textcolor{black}{and }at distances of up to $h\sim10^3\lambda_\text{ENZ}$. Thus, the time-modulated plasmonic slab acts as an antenna \textcolor{black}{at $\omega\sim0.5\Omega$,} radiating electromagnetic energy to the far-field and modifying the emission of light at distances much larger than in the static case. These oscillations become compressed as $h$ increases and eventually die out for $h\sim5\cdot10^4\lambda_\text{ENZ}$ (see the inset), where the free-space value is recovered.
\par
Lastly, commenting on the values chosen, the specific value of $\alpha$ is not important as long as it is above the threshold of losses of the system $\gamma$ \footnote{The threshold of losses of the system is determined by the damping rate $\gamma$ and the larger the damping rate, the larger this threshold becomes.}. On the other hand, having $\Omega\gtrsim2\omega_{\text{SP}}$ allows the electric dipole to efficiently couple to $\Omega - \omega_{\text{SP}}$, the negative frequency replica of the surface plasmon, which then lies close to the light-cone. \textcolor{black}{Additionally, $\omega_\text{B}$ and $\Omega - \omega_\text{B}$ always interact close to the parametric resonance condition, $\omega\sim0.5\Omega$. Having $\Omega\simeq2\omega_{\text{ENZ}}$ then makes the two modes interact more strongly for a given $\alpha$, since in that case the interaction point lies at $\omega\sim\omega_\text{ENZ}$ and $K_\parallel\sim0$, where both modes are flat \cite{supmat}.}
\par
\textit{Conclusions---}We have demonstrated that temporal modulation of a plasmonic slab can fundamentally alter dipole emission, transforming near-field quenching into non-radiative gain and enabling far-field radiation control at unprecedented distances. In particular, coupling to the negative-frequency replica of the surface plasmon produces a flat gain band that turns a dipole into an energy absorber, while \textcolor{black}{the emergence of a leaky resonance associated with the Brewster condition and its negative frequency replica} yields large, long-lived far-field oscillations reaching the free-space emission level up to $10^{3}\lambda_\text{ENZ}$. These results establish a new route for tailoring light–matter interaction in time-varying plasmonic systems by simultaneously exploiting spatial confinement and material dispersion. Beyond the classical regime studied here, our findings suggest analogous phenomena for quantum emitters, opening opportunities for persistent excitation and emission engineering in plasmonic time crystals.

\begin{acknowledgments}
\textit{Acknowledgments---}We acknowledge funding from the European Union through the ERC grant TIMELIGHT under GA101115792. PAH acknowledges support from the Spanish Ministry of Science and Innovation through the Ramón y Cajal program (Grant No. RYC2021-031568-I) and through Project No. PID2022-141036NA-I00. JESO also acknowledges support from the CAM Consejería de Educación, Ciencia y Universidades, Viceconsejería de Universidades, Investigación y Ciencia, Dirección General de Investigación e Innovación Tecnológica (CAM FPI grant Ref. A281).

\end{acknowledgments}

\bibliography{main}% Produces the bibliography via BibTeX.

\end{document}

\maketitle
\thispagestyle{empty} % Remove page number on the first page
\setstcolor{red}

\vspace*{\fill} % Empujar el contenido hacia abajo

\vspace*{\fill} % Empujar el contenido hacia arriba

\tableofcontents

\newpage
% Set numbering for supplemental material
\pagenumbering{arabic} % Reset numbering
\setcounter{page}{1} % Start from 1
\renewcommand{\thepage}{S\arabic{page}} % Prefix 'S' before the page number

\section{Scattering Matrix}

We consider the optical response of a time modulated plasmonic slab, such as ITO. Taking its response to be dominated by conduction band electrons, the system obeys the following driven-Drude differential equation

\begin{equation}
    \partial^2_{t}\textbf{P} + \gamma\textbf{P} = \varepsilon_0\omega_\text{p}^2(t)\textbf{E},
\end{equation}

where $\textbf{E}$ is the external electric field acting on the electrons, and where

\begin{equation}
    \omega_\text{p}^2(t) = \omega_\text{p}^2[1 + \alpha\cos(\Omega t)].
\end{equation}

In the above, $\omega_\text{p}$ is the static plasma frequency, whereas $\alpha$ and $\Omega$ are, respectively, the modulation strength and frequency. Given the periodicity of the plasma frequency, we can make use of Floquet's theorem and work with the following ansatzs for the polarization and the electric field: $\text{P}_{j}(t) = \sum_ne^{-i\omega_nt}\text{P}_{j,n}$ and $\text{E}_{j}(t) = \sum_ne^{-i\omega_nt}\text{E}_{j,n}$, where $j=x\text{, }y\text{, z}$ labels the different Cartesian components, and $\omega_n = \omega + n\Omega$, where $\omega$ is real-valued, and $n\in\mathbb{Z}$. Under these Floquet ansatzs, we can explicitly solve the inhomogeneous differential equation as

\begin{equation}
    \text{P}_{\alpha,n} = \varepsilon_0\sum_{n'}\chi_{n,n'}\text{E}_{\alpha,n'},
\end{equation}

where we introduce the following Floquet susceptibility matrix:

\begin{equation}
    \begin{split}
        \chi_{n,n'} & = \Big(\varepsilon_\infty - 1 - \frac{\omega_\text{p}^2}{\omega_n^2 + i\gamma\omega_n}\Big)\delta_{n,n'} -
        \frac{\alpha}{2}\frac{\omega_\text{p}^2}{\omega_n^2 + i\gamma\omega_n}\big(\delta_{n+1,n'} + \delta_{n-1,n'}\big).
    \end{split}
\end{equation}

The Floquet susceptibility matrix as defined above already includes the $\omega\longrightarrow\infty$ optical response of the system, which is not captured by the driven-Drude equation. Additionally, we can see that the material response is both dispersive $-$owing to the frequency dependence$-$ and time-modulated, since the frequency domain Floquet susceptibility is not a scalar, but a matrix, which allows different harmonics to interact. An important feature of the Floquet susceptibility matrix is that it connects positive and negative frequencies. It is well-known that negative frequency fields are nothing but complex conjugates of positive frequency ones (for the case of real frequencies). Importantly, the emergence of gain in our system is rooted precisely in the coupling of positive frequency modes to negative frequency ones.
\\
Having discussed the optical response of the plasmonic time-varying slab, we turn our attention to Maxwell curl equations

\begin{subequations}
    \begin{equation}
        \curl\textbf{E} = -\partial_t\textbf{B},
    \end{equation}
    \begin{equation}
        \curl\textbf{H} = \partial_t\textbf{D}.
    \end{equation}
\end{subequations}

To proceed any further, we need to make use of the constitutive relations between the different fields. In Fourier space, and under Floquet ansatzs for the fields, these relations read $\text{B}_{j,n} = \mu_0\text{H}_{j,n}$ and $\text{D}_{j,n} = \varepsilon_0\sum_{n'}\varepsilon_{n,n'}\text{E}_{j,n'}$, where we have introduced a Floquet permittivity matrix $\varepsilon_{n,n'} = \delta_{n,n'} + \chi_{n,n'}$.
\\
On the other hand, due to the planar boundaries of the slab, the lateral momentum $\boldsymbol{K}_\parallel$ is a conserved quantity, as it is the polarization $\sigma =$TE, TM of the fields. Additionally, since the system has planar symmetry \footnote{By planar symmetry we mean that the system remains invariant under arbitrary rotations around the direction normal to the air-metal interfaces.}, we can without loss of generality assume $\boldsymbol{K}_\parallel = K_\parallel\boldsymbol{e}_y$ and furthermore, that $K_\parallel\geq0$. Hence, inside the metal, we look for solutions of the form:
\begin{equation}
    \text{E}^{(m)}_j(y,z,t) = e^{i(k^{(m)}_zz + K_\parallel y)}\sum_n\text{E}^{(m)}_{j,n}e^{-i\omega_nt},
\end{equation}
and
\begin{equation}
    \text{H}^{(m)}_j(y,z,t) = e^{i(k^{(m)}_zz + K_\parallel y)}\sum_n\text{H}^{(m)}_{j,n}e^{-i\omega_nt},
\end{equation}

where $k_z^{(m)} = k_z^{(m)}(\omega, K_\parallel)$ is the normal momentum inside the time-modulated plasmonic slab, whose dependence on both $\omega$ and $K_\parallel$ shall be given by the corresponding dispersion relation.
\\
Similarly, we write the electric and magnetic fields outside the slab as:
\begin{equation}
    \text{E}^\text{(L)}_{j}(y,z,t) = e^{iK_\parallel y}\sum_n\text{E}^\text{(in)}_{f,j,n}e^{(k_{z,n}z-i\omega_nt)} + e^{iK_\parallel y}\sum_n\text{E}^\text{(out)}_{b,j,n}e^{(-k_{z,n}z-i\omega_nt)},
\end{equation}
\begin{equation}
    \text{H}^\text{(L)}_{j}(y,z,t) = e^{iK_\parallel y}\sum_n\text{H}^\text{(in)}_{f,j,n}e^{(k_{z,n}z-i\omega_nt)} + e^{iK_\parallel y}\sum_n\text{H}^\text{(out)}_{b,j,n}e^{(-k_{z,n}z-i\omega_nt)},
\end{equation}
at its left side and
\begin{equation}
    \text{E}^\text{(R)}_{j}(y,z,t) = e^{iK_\parallel y}\sum_n\text{E}^\text{(out)}_{f,j,n}e^{(k_{z,n}z-i\omega_nt)} + e^{iK_\parallel y}\sum_n\text{E}^\text{(in)}_{b,j,n}e^{(-k_{z,n}z-i\omega_nt)},
\end{equation}
\begin{equation}
    \text{H}^\text{(R)}_{j}(y,z,t) = e^{iK_\parallel y}\sum_n\text{H}^\text{(out)}_{f,j,n}e^{(k_{z,n}z-i\omega_nt)} + e^{iK_\parallel y}\sum_n\text{H}^\text{(in)}_{b,j,n}e^{(-k_{z,n}z-i\omega_nt)},
\end{equation}
at its right one.\\
In the above formulae, $f$ stands for \textit{forward} (field propagating towards $z>0$), $b$ stands for \textit{backward} (field propagating towards $z<0$) and $\text{in}$ and $\text{out}$, for \textit{input} and \textit{output} fields, respectively. Also, $k_{z,n} = \theta(\abs{\omega_n} - c_0K_\parallel)\text{sgn}(\omega_n)\sqrt{(\omega_n/c_0)^2 - K^2_\parallel} \text{ }+\text{ } \theta(c_0K_\parallel - \abs{\omega_n})i\sqrt{K^2_\parallel - (\omega_n/c_0)^2}$; the inclusion of $\text{sgn}(\omega_n)$ for the case of real wave-vector is necessary to have negative frequency waves be complex conjugates of positive frequency ones.
\\
We now solve separately for TE and TM-polarized fields. In the former case, the electric field is perpendicular to the plane of incidence, so that $\text{E}_{z,n} = 0$, whereas in the latter, it is the magnetic field that is normal to the plane of incidence, with $\text{H}_{z,n} = 0$. For TE-polarized fields, Maxwell equations read:
\begin{subequations}
    \begin{equation}
        \mu_0\omega_n\text{H}^{(m)}_{z,n} = -K_\parallel \text{E}_{x,n}^{(m)},
    \end{equation}
     \begin{equation}
        \mu_0\omega_n\text{H}^{(m)}_{y,n} = k^{(m)}_z\text{E}_{x,n}^{(m)},
    \end{equation}
     \begin{equation}
        k_z\text{H}^{(m)}_{y,n} - K_\parallel\text{H}^{(m)}_{z,n} = \varepsilon_0\omega_n\sum_{n'}\varepsilon_{n,n'}\text{E}^{(m)}_{x,n'}.
    \end{equation}
\end{subequations}

We now define the following matrices:
\begin{equation}
    \text{M}^\text{HE}_{n,n'} = \varepsilon_0\omega_{n}\varepsilon_{n,n'} - \frac{K_\parallel^2}{\mu_0\omega_n}\delta_{n,n'},
\end{equation}
\begin{equation}
    \text{M}^\text{EH}_{n,n'} = \mu_0\omega_{n}\delta_{n,n'},
\end{equation}

which enable us to cast the following eigenvalue problem for $k^{(m)}_x$ and the Floquet amplitudes $\text{E}^{(m)}_{z,n}$:
\begin{equation}
    (k^{(m)}_x)^2\textbf{E}^{(m)}_{z} = \textbf{M}^\text{EH}\cdot\textbf{M}^\text{HE}\cdot\textbf{E}^{(m)}_{z},
\end{equation}
where 
\begin{equation}
    (\textbf{M}^\text{EH}\cdot\textbf{M}^\text{HE})_{n,n'} = \frac{\omega^2_{n}}{c_0^2}\varepsilon_{n,n'} - K_\parallel^2\delta_{n,n'}.
\end{equation}
For practical purposes, we have to set a bound on the number of Floquet modes included in the field expansions. Therefore, we take $-N_\text{F}\leq n\leq N_\text{F}$, where $2N_\text{F} + 1$ is the total number of modes included in the expansion. We also mention that the eigenvalue problem is defined for $(k^{(m)}_z)^2$ and so, both $k^{(m)}_z$ and $-k^{(m)}_z$ need to be taken into account when writing the total electric and magnetic fields inside the time-varying plasmonic slab. Thus, we have a total of $2\cdot(2N_\text{F} + 1)$ different normal wave-vectors $k^{(m)}_{z,p}$ inside the metal, where $-2N_\text{F}\leq p\leq2N_\text{F}$, and where $k^{(m)}_{z,-p} = -k^{(m)}_{z,p}$.
\\
We now turn our attention to the TM-modes, for which we have:
\begin{subequations}
    \begin{equation}
        \mu_0\omega_n\text{H}^{(m)}_{x,n} = -k^{(m)}_{z}\text{E}^{(m)}_{y,n} + K_\parallel \text{E}^{(m)}_{z,n},
    \end{equation}
    \begin{equation}
        K_\parallel \text{H}^{(m)}_{x,n} =  \varepsilon_0\omega_n\sum_{n'}\varepsilon_{n,n'}\text{E}^{(m)}_{z,n'},
    \end{equation}
    \begin{equation}
        k^{(m)}_z\text{H}^{(m)}_{x,n} = -\varepsilon_0\omega_{n}\sum_{n'}\varepsilon_{n,n'}\text{E}^{(m)}_{y,n'}.
    \end{equation}
\end{subequations}

We introduce a set of matrices for TM-polarized waves:

\begin{equation}
    \text{M}^{\text{HE}}_{n,n'} = -\varepsilon_0\omega_n\varepsilon_{n,n'} ,
\end{equation}
\begin{equation}
    \text{M}^\text{EH}_{n,n'} = -\mu_0\omega_n\delta_{n,n'} +K^2_\parallel[(\text{M}^\text{HE})^{-1}]_{n,n'},
\end{equation}

which enable us to write the TM-polarization corresponding eigenvalue problem for $k^{(m)}_z$ and the Floquet amplitudes $\text{E}^{(m)}_{x,n}$:
\begin{equation}
    (k^{(m)}_z)^2\textbf{E}^{(m)}_{x} = \textbf{M}^\text{EH}\cdot\text{M}^\text{HE}\cdot\textbf{E}^{(m)}_{x},
\end{equation}
where 
\begin{equation}
    (\textbf{M}^\text{EH}\cdot\textbf{M}^\text{HE})_{n,n'} = \frac{\omega^2_{n}}{c_0^2}\varepsilon_{n,n'} - K_\parallel^2\delta_{n,n'}.
\end{equation}
Thus, the eigenvalue problem is the same for both polarizations, as it should, since the eigenvalue problem is defined for the bulk of the time-modulated metal.
\\
Taking into account all the different modes $k^{(m)}_{z,p}$, the tangential components of the electric and magnetic fields inside the time-modulated ITO slab read:
\begin{equation}
    \text{E}^{(m)}_{\sigma,t} = \sum_{p}\sum_{n}e^{i(k^{(m)}_{z,p}z + K_\parallel y)}e^{-i\omega_nt}\mathcal{E}^{(m)}_{\sigma,p,n}\text{E}^{(m)}_{\sigma,p},
\end{equation}
and
\begin{equation}
    \text{H}^{(m)}_{\sigma,t} = \sum_{p}\sum_{n}e^{i(k^{(m)}_{z,p}z + K_\parallel y)}e^{-i\omega_nt}\mathcal{H}^{(m)}_{\sigma,p,n}\text{H}^{(m)}_{\sigma,p}.
\end{equation}
In the above, $\mathcal{E}^{(m)}_{\sigma,p,n}$ and $\mathcal{H}^{(m)}_{\sigma,p,n}$ are the amplitudes of the $n$-th Floquet harmonic component of the $p$-th eigenvector inside the metal, for the case of $\sigma$-polarized waves. Hence, $p$ labels the eigenvector, while $n$ labels the Floquet harmonic. Additionally, the $\mathcal{E}^{(m)}_{\sigma,p,n}$ are given by the eigenvalue problem inside the plasmonic time crystal, while $\mathcal{H}^{(m)}_{\sigma,p,n} = \sum_{n'}\text{M}^\text{HE}_{\sigma;n,n'}\mathcal{E}^{(m)}_{\sigma,p,n}/k^{(m)}_{z,p}$, where now we have made explicit the dependence of $\text{M}^\text{HE}_{\sigma;n,n'}$ on the polarization. \footnote{The $\mathcal{E}^{(m)}_{\sigma,p,n}$ are in truth independent of $\sigma$, whereas the $\mathcal{H}^{(m)}_{p,n}$ are different for TE and TM modes.} Lastly, $\text{E}^{(m)}_{\sigma,p}$ is the amplitude of the $p$-th eigenvector for $\sigma$ polarization. The latter is not given by the eigenvalue problem, but by the scattering matrix, which we derive now.
\\
Imposing the continuity of the tangential components of the electric and magnetic field on both the left and right air-metal interfaces, we have:
\begin{subequations}
    \begin{equation}
        \text{E}^\text{(in)}_{\sigma,f,n} + \text{E}^\text{(out)}_{\sigma,b,n} = \sum_{p}\mathcal{E}^{(m)}_{\sigma,p,n}\text{E}^{(m)}_{\sigma,p},
    \end{equation}
    \begin{equation}
        \text{H}^\text{(in)}_{\sigma,f,n} + \text{H}^\text{(out)}_{\sigma,b,n} = \sum_{p}\mathcal{H}^{(m)}_{\sigma,p,n}\text{E}^{(m)}_{\sigma,p},
    \end{equation}
\end{subequations}
to the left, and
\begin{subequations}
    \begin{equation}
        \text{E}^\text{(out)}_{\sigma,f,n} + \text{E}^\text{(in)}_{\sigma,b,n} = \sum_{p}\mathcal{E}^{(m)}_{\sigma,p,n}\text{E}^{(m)}_{\sigma,p}e^{ik^{(m)}_{z,p}d},
    \end{equation}
    \begin{equation}
        \text{H}^\text{(out)}_{\sigma,f,n} + \text{H}^\text{(in)}_{\sigma,b,n} = \sum_{p}\mathcal{H}^{(m)}_{\sigma,p,n}\text{E}^{(m)}_{\sigma,p}e^{ik^{(m)}_{z,p}d},
    \end{equation}
\end{subequations}
to the right.
\\
To go any further, we have to relate the magnetic field to the electric one. In free-space, we have $\text{H}_{\sigma,f,n} = \text{Y}_{\sigma,n}\text{E}_{\sigma,f,n}$, where $\text{Y}_{\text{TE},n} = -k_{z,n} /( \mu_0\omega_n)$ and $\text{Y}_{\text{TM},n}= \varepsilon_0\omega_n/ k_{z,n}$. $\text{Y}_{\sigma,n}$ is actually the mode admittance for $\sigma$-polarized waves with frequency $\omega_n$ and lateral momentum $K_\parallel$. For backward modes, the relations flip sign, and we ought to use $\text{Y}_{\text{TE},n}\longrightarrow k_{z,n} /( \mu_0\omega_n)$ and $\text{Y}_{\text{TM},n} \longrightarrow-\varepsilon_0\omega_n/ k_{z,n}$. Likewise, inside the time-varying metal, we can use $\mathcal{H}^{(m)}_{\sigma,p,n} = \sum_{n'}\text{M}^\text{HE}_{\sigma;n,n'}\mathcal{E}^{(m)}_{\sigma,p,n}/k^{(m)}_{z,p}$. Thus, the continuity of the tangential component of the magnetic field reads

\begin{equation}
        -\text{E}^\text{(in)}_{\sigma,f,n} + \text{E}^\text{(out)}_{\sigma,b,n} = -\frac{1}{\text{Y}_{\sigma,n}}\sum_{p}\left(\sum_{n'}\text{M}^\text{HE}_{\sigma;n,n'}\mathcal{E}^{(m)}_{\sigma,p,n'}/k^{(m)}_{z,p}\right)\text{E}^{(m)}_{\sigma,p}
\end{equation}

at the left air-metal interface, and

\begin{equation}
        \text{E}^\text{(out)}_{\sigma,f,n} - \text{E}^\text{(in)}_{\sigma,b,n} = \frac{1}{\text{Y}_{\sigma,n}}\sum_{p}\left(\sum_{n'}\text{M}^\text{HE}_{\sigma;n,n'}\mathcal{E}^{(m)}_{\sigma,p,n'}/k^{(m)}_{z,p}\right)\text{E}^{(m)}_{\sigma,p}e^{ik^{(m)}_{z,p}d}
\end{equation}
at the right one.
\\
Next, we write $\mathcal{E}^{(m)}_{\sigma,p,n'}$ as a rectangular $(2N_\text{F} + 1)\cross2(2N_\text{F} + 1)$ matrix $\boldsymbol{\mathcal{E}}^\text{L}_{\sigma}$, with elements $[\boldsymbol{\mathcal{E}}^\text{L}_{\sigma}]_{n,p} = \mathcal{E}^{(m)}_{\sigma,p,n}$. Likewise, we define $[\boldsymbol{\mathcal{E}}^\text{R}_{\sigma}]_{n,p} = \mathcal{E}^{(m)}_{\sigma,p,n}e^{ik^{(m)}_{z,p}d}$, as well as $[\boldsymbol{\mathcal{H}}^\text{L}_{\sigma}]_{n,p} = (-1/\text{Y}_{\sigma,n})\sum_{n'}\text{M}^\text{HE}_{\sigma;n,n'}\mathcal{E}^{(m)}_{\sigma,p,n'}/k^{(m)}_{z,p}$ and $[\boldsymbol{\mathcal{H}}^\text{R}_{\sigma}]_{n,p} = (1/\text{Y}_{\sigma,n})\sum_{n'}\left(\text{M}^\text{HE}_{\sigma;n,n'}\mathcal{E}^{(m)}_{\sigma,p,n'}/k^{(m)}_{z,p}\right)e^{ik^{(m)}_{z,p}d}$, where in the definition of the $\boldsymbol{\mathcal{H}}^\text{L, R}_\sigma$ we have included the $\text{Y}_{\sigma,n}$. We also write the different Floquet/eigenvector components in vector form, with $[\textbf{E}^\text{(in)}_{\sigma,f}]_n = \text{E}^\text{(in)}_{\sigma,f,n}$ and $[\textbf{E}^{(m)}_{\sigma}]_p = \text{E}^{(m)}_{\sigma,p}$ and similarly for the others, where we recall that $-N_\text{F}\leq n\leq N_\text{F}$ and $-2N_\text{F}\leq p \leq 2N_\text{F}$. In doing so, the continuity of the tangential electric and magnetic fields reads

\begin{equation}
    \begin{bmatrix}
        \text{}\boldsymbol{\text{I}} && \boldsymbol{\text{I}} \\ \\ -\boldsymbol{\text{I}} && \boldsymbol{\text{I}}
    \end{bmatrix}\cdot\begin{bmatrix}
        \textbf{E}^\text{(in)}_{\sigma,f} \\ \\\textbf{E}^\text{(out)}_{\sigma,b}
    \end{bmatrix} = \begin{bmatrix}
        \boldsymbol{\mathcal{E}}^\text{L}_{\sigma} \\ \\\boldsymbol{\mathcal{H}}^\text{L}_{\sigma}
    \end{bmatrix}\cdot\textbf{E}^{(m)}_\sigma = \boldsymbol{\mathcal{M}}^\text{L}_{\sigma}\cdot\textbf{E}^{(m)}_\sigma,
\end{equation}
and

\begin{equation}
    \begin{bmatrix}
        \text{}\boldsymbol{\text{I}} && \boldsymbol{\text{I}} \\ \\ \boldsymbol{\text{I}} && -\boldsymbol{\text{I}}
    \end{bmatrix}\cdot\begin{bmatrix}
        \textbf{E}^\text{(out)}_{\sigma,f} \\ \\ \textbf{E}^\text{(in)}_{\sigma,b}
    \end{bmatrix} = \begin{bmatrix}
        \boldsymbol{\mathcal{E}}^R_{\sigma} \\ \\\boldsymbol{\mathcal{H}}^R_{\sigma}
    \end{bmatrix}\cdot\textbf{E}^{(m)}_\sigma = \boldsymbol{\mathcal{M}}^\text{R}_{\sigma}\cdot\textbf{E}^{(m)}_\sigma,
    \label{eq:eR_eq}
\end{equation}

where $\boldsymbol{\text{I}}$ is the $(2N_\text{F}+1)\cross (2N_\text{F}+1)$ identity matrix and $\boldsymbol{\mathcal{M}}^\text{L}_{\sigma}$ and $\boldsymbol{\mathcal{M}}^\text{R}_{\sigma}$ are both square matrices of rank $2(2N_\text{F} + 1)$. We can invert Eq.~\eqref{eq:eR_eq}, with

\begin{equation}
    \textbf{E}^{(m)}_\sigma = \left(\boldsymbol{\mathcal{M}}^\text{L}_{\sigma}\right)^{-1}\cdot\begin{bmatrix}
        \text{}\boldsymbol{\text{I}} && \boldsymbol{\text{I}} \\ \\ -\boldsymbol{\text{I}} && \boldsymbol{\text{I}}
    \end{bmatrix}\cdot\begin{bmatrix}
        \textbf{E}^\text{(in)}_{\sigma,f} \\ \\\textbf{E}^\text{(out)}_{\sigma,b}
    \end{bmatrix}.
\end{equation}
and now substitute it back in equation (28). Doing so, and rearranging the terms involved, we arrive to:

\begin{equation}
    \begin{bmatrix}
        \begin{bmatrix}
            \boldsymbol{\text{ I }} \\ \\ \boldsymbol{\text{ I }}
        \end{bmatrix} &, -\boldsymbol{\mathcal{M}}^\text{R}_{\sigma}\cdot\begin{bmatrix}
            \boldsymbol{\text{ I }} \\ \\ \boldsymbol{\text{ I }}
        \end{bmatrix}
    \end{bmatrix}\cdot\begin{bmatrix}
        \textbf{E}^\text{(out)}_{\sigma,f} \\ \\ \textbf{E}^\text{(out)}_{\sigma,b}
    \end{bmatrix} = \begin{bmatrix}
        \boldsymbol{\mathcal{M}}^\text{R}_{\sigma}\cdot\left(\boldsymbol{\mathcal{M}}^\text{L}_{\sigma}\right)^{-1}\cdot\begin{bmatrix}
            \boldsymbol{\text{ I }} \\ \\ -\boldsymbol{\text{ I }}
        \end{bmatrix} &, \begin{bmatrix}
            -\boldsymbol{\text{ I }} \\ \\ \boldsymbol{\text{ I }}
        \end{bmatrix}
    \end{bmatrix}\cdot\begin{bmatrix}
        \textbf{E}^\text{(in)}_{\sigma,f} \\ \\ \textbf{E}^\text{(in)}_{\sigma,b}
    \end{bmatrix}.
\end{equation}

Calling the left-hand side matrix $\boldsymbol{S}^\text{(out)}_{\sigma}$ and the right-hand one $\boldsymbol{S}^\text{(in)}_{\sigma}$, the scattering matrix is given by 

\begin{equation}
    \boldsymbol{S}_{\sigma}(\omega,K_\parallel) = \left(\boldsymbol{S}^\text{(out)}_{\sigma}\right)^{-1}\cdot\boldsymbol{S}^\text{(in)}_{\sigma},
\end{equation}

where we have made explicit the dependence in both $\omega$ and $K_\parallel$, which we had so far omitted for the sake of simplicity.
\\
The scattering matrix can be written as follows:

\begin{equation}
    \boldsymbol{S}_{\sigma}(\omega,K_\parallel) = \begin{bmatrix}
        \boldsymbol{T}_{\sigma}(\omega,K_\parallel) && \boldsymbol{R}_{\sigma}(\omega,K_\parallel) \\ \\ \boldsymbol{R}'_{\sigma}(\omega,K_\parallel) && \boldsymbol{T}'_{\sigma}(\omega,K_\parallel)
    \end{bmatrix}.
    \label{eq:s-matrix2}
\end{equation}

In Eq.~\eqref{eq:s-matrix2}, we have introduced transmission and reflection matrices, whose elements $[\boldsymbol{T}_\sigma(\omega,K_\parallel)]_{n,n'} = T_\sigma(\omega_n,\omega_{n'};K_\parallel)]$ and $[\boldsymbol{R}_\sigma(\omega,K_\parallel)]_{n,n'} = R_\sigma(\omega_n,\omega_{n'};K_\parallel)]$ are generalized Fresnel-Floquet transmission and reflection coefficients that connect an incident frequency $\omega_{n'}$ with an outgoing one $\omega_n$.
\\
It is possible to find more compact expressions for the transmission and reflection matrices. Let $[\boldsymbol{\mathcal{E}}^{(+)}_{\sigma}]_{n,p} = \mathcal{E}^{(m)}_{\sigma,p,n}$ with $0 \leq p\leq2N_\text{F}$ (recall that, originally, $-2N_\text{F} \leq p\leq2N_\text{F}$). Let us as well define $[\boldsymbol{\mathcal{H}}^{(+)}_{\sigma}]_{n,p} = (-1/\text{Y}_{\sigma,n})\sum_{n'}\text{M}^\text{HE}_{\sigma;n,n'}\mathcal{E}^{(m)}_{\sigma,p,n'}/k^{(m)}_{z,p}$, where once again $0 \leq p\leq2N_\text{F}$. Hence, we see $\boldsymbol{\mathcal{E}}^{(+)}_{\sigma}$ and $\boldsymbol{\mathcal{H}}^{(+)}_{\sigma}$ are square matrices of rank $2N_\text{F} + 1$. Now, we introduce diagonal cosine and sine matrices, $[\textbf{C}]_{p,p'} = \text{cos}(k^{(m)}_{z,p}d)\delta_{p,p'}$ and $[\textbf{S}]_{p,p'} = \text{sin}(k^{(m)}_{z,p}d)\delta_{p,p'}$, where $0\leq p,p'\leq 2N_\text{F}$. Next, let us define matrices $\textbf{A}_\sigma$ and $\textbf{B}_\sigma$, where
\begin{equation}
    \textbf{A}_\sigma = \boldsymbol{\mathcal{H}}^{(+)}_{\sigma}\cdot\big[i\textbf{S}\cdot\big(\boldsymbol{\mathcal{E}}^{(+)}_{\sigma}\big)^{-1} + \textbf{C}\cdot\big(\boldsymbol{\mathcal{H}}^{(+)}_{\sigma}\big)^{-1}\big]
\end{equation}
and
\begin{equation}
    \textbf{B}_\sigma = \boldsymbol{\mathcal{E}}^{(+)}_{\sigma}\cdot\big[i\textbf{S}\cdot\big(\boldsymbol{\mathcal{H}}^{(+)}_{\sigma}\big)^{-1} + \textbf{C}\cdot\big(\boldsymbol{\mathcal{E}}^{(+)}_{\sigma}\big)^{-1}\big].
\end{equation}
In terms of these matrices, the transmission and reflection matrices read:
\begin{equation}
    \boldsymbol{T}_\sigma = 2(\textbf{A}_\sigma + \textbf{B}_\sigma)^{-1}
\end{equation}
and
\begin{equation}
    \boldsymbol{R}_\sigma = -\textbf{I} + \textbf{B}_\sigma\cdot\boldsymbol{T}_\sigma.
\end{equation}
\textcolor{black}{In all the calculations we have used $N_\text{F} = 5$, which ensures numerical convergence. Therefore, a total of $2N_\text{F} + 1 = 11$ bands have been included.}

\section{Dipole Radiation}

We now consider a harmonic electric point dipole, oscillating at frequency $\omega_{\text{d}}$ at a height $h$ above the air-metal interface, outside the time-modulated slab. The current associated to the electric dipole is given by

\begin{equation}
    \textbf{J}_{\text{d}}(\textbf{x},t) = -i\omega_{\text{d}}\textbf{d}e^{-i\omega_{\text{d}}t}\delta(\textbf{x} - \textbf{x}_0) + \text{c.c.},
\end{equation}

where $\textbf{d}$ is its dipole moment, with components $d_x,\text{ }d_y\text{ and }d_z$.
\\
When placed in front of the metallic slab, the dipole interacts with its own field and the one reflected from the surface. Following \cite{novotny2012principles}, the electric field of the dipole reads

\begin{equation}
    \begin{split}
        \textbf{E}^\text{d}(\textbf{x},t) = \sum_{\sigma = \text{TE, TM}}\int\frac{d^2K_\parallel}{(2\pi)^2}\frac{1}{k_z}e^{-i\omega_{\text{d}}t}e^{i(\boldsymbol{K}_\parallel\cdot\textbf{x}_\parallel +k_z\abs{z + h})}\text{E}_\sigma(\omega_\text{d},\boldsymbol{K}_\parallel)\textbf{e}^{\pm}_\sigma(\omega_\text{d},\boldsymbol{K}_\parallel) + \text{c.c},
    \end{split}
\end{equation}

where 

\begin{equation}
    \text{E}^{\pm}_\sigma(\omega_\text{d},\boldsymbol{K}_\parallel) = \frac{\omega_\text{d}^2}{\varepsilon_0c_0^2}\textbf{d}\cdot\textbf{e}^\pm_\sigma(\omega_\text{d},\boldsymbol{K}_\parallel),
    \label{eq:e-field_dipole}
\end{equation}

where $\textbf{e}^\pm_\text{TE}(\omega_\text{d},\boldsymbol{K}_\parallel) = \textbf{e}_\text{TE}(\omega_\text{d},\boldsymbol{K}_\parallel) = \frac{1}{K_\parallel}(k_y,\text{ }k_z,0)^\text{T}$ and $\textbf{e}^{\pm}_\text{TM}(\omega_\text{d},\boldsymbol{K}_\parallel) = \frac{c_0}{\omega_\text{d}K_\parallel}(\pm k_xk_z,\text{ }\mp k_xk_y,\text{ } K_\parallel^2)^\text{T}$. In Eq.~\eqref{eq:e-field_dipole}, $k_z = \sqrt{(\omega_\text{d} / c_0)^2 - K_\parallel^2}$, while $\boldsymbol{K}_\parallel = (k_x,\text{ }k_y,0)^\text{T}$ and $\textbf{x}_\parallel = (x,\text{ }y,0)^\text{T}$. On the other hand, the $\pm$ sign on top of $\textbf{e}^{\pm}_\text{TM}(\omega_\text{d},\boldsymbol{K}_\parallel)$ is that of $z + h$; hence, we have $+$ for $z + h > 0$ and $-$ if $z + h < 0$. According to our choice of coordinates, the dipole is at $z = -h$, while the left air-metal interface is at $z = 0$.
\\
The field reflected from the surface shall contain many (in theory, infinitely many) harmonics $\omega_\text{d} + n\Omega$, so that the reflected field reads

\begin{equation}
    \begin{split}
        \textbf{E}^{\text{ref}}(\textbf{x},t) & = \sum_{n,\sigma}e^{-i(\omega_{\text{d}} + n\Omega)t}\int\frac{d^2K_\parallel}{(2\pi)^2}\frac{1}{k_{n,z}}e^{i(\boldsymbol{K}_\parallel\cdot\boldsymbol{x}_\parallel -k_{n,z}z)}R_\sigma(\omega_\text{d} + n\Omega,\omega_\text{d})\text{E}^{+}_\sigma(\omega_\text{d},\boldsymbol{K}_\parallel)e^{ik_zh}\textbf{e}^-_\sigma(\omega_\text{d},\boldsymbol{K}_\parallel) \\& + \text{c.c.},
    \end{split}
\end{equation}

where $k_{n,z} = \sqrt{(\omega_\text{d} + n\Omega)^2/ c_0^2 - K_\parallel^2}$.
\\
Now, the total energy exchanged per unit time between the dipole and the total electric field (which includes its own and the one reflected back from the surface) is

\begin{equation}
    P(t) = \int d^3x\textbf{J}_\text{d}\cdot(\textbf{E}^{\text{d}} + \textbf{E}^\text{ref}).
\end{equation}

We now integrate in time from $t' = t_0$ to $t' = t$, divide by $t - t_0$ and take the $t\longrightarrow\infty$ limit. We then make use of the following identities:

\begin{equation}
    \underset{t\longrightarrow\infty}{\lim}\frac{\sin\left(n\Omega(t - t_0)\right)}{n\Omega(t - t_0)} = \delta_{n,0},
\end{equation}

and 

\begin{equation}
    \underset{t\longrightarrow\infty}{\lim}\frac{\sin[\left(2\omega_{\text{d}} + n\Omega\right)(t - t_0)]}{\left(2\omega_{\text{d}} + n\Omega\right)(t - t_0)} = 0.
\end{equation}

%Although the above is not technically zero when $\omega_\text{d} = 0.5\Omega$ and $n=-1$, for $\omega_\text{d} = 0.5\Omega$, $e^{-i(\omega_\text{d} - \Omega)t} = e^{i\omega_\text{d}t}$ is already taken into account in the complex conjugate $\text{c.c.}$ and so, we have to take the second limit to be zero or otherwise, we would be counting twice the corresponding harmonic.
The above limit is not zero when $\omega_\text{d} = 0.5\Omega$ and $n=-1$, at exactly the parametric amplifier condition. However, since (realistically) any dipole can be infinitely close to, but not exactly at $\omega_\text{d} = 0.5\Omega$, we can always take the limit to vanish.
\\
Bearing this in mind, we see that the dipole couples only to the $n = 0$ harmonic, the fundamental one. Therefore, even though the different $\omega_\text{d} + n\Omega$ have a non-vanishing contribution to the transient, they do not contribute to the total power $P_\text{tot} = \underset{t\longrightarrow\infty}{\lim}\frac{1}{t - t_0}\int_{t_0}^tP(t)$. Therefore, the formula for the total power emitted by the dipole reads \cite{novotny2012principles}

\begin{equation}
    \begin{split}
        \frac{P_\text{tot}}{P_0} & = 1 + \frac{d_x^2 + d_y^2}{d^2}\frac{3\omega_\text{d}^2}{4c_0^2}\int_0^\infty d K_\parallel\mathfrak{R}\left\{\frac{K_\parallel}{k_z}\Big(R_\text{TE}(\omega_\text{d},\omega_\text{d};K_\parallel) - k_z^2R_\text{TM}(\omega_\text{d},\omega_\text{d};K_\parallel)\Big)e^{i2k_zh}\right\} \\&
        +\frac{d_z^2}{d^2}\frac{3\omega_\text{d}^2}{2c_0^2}\int_0^\infty dK_\parallel\mathfrak{R}\left\{\frac{K_\parallel^3}{k_z}R_\text{TM}(\omega_\text{d},\omega_\text{d};K_\parallel)e^{i2k_zh}\right\},
    \end{split}
\end{equation}

where we have taken the frequency conserving Fresnel-Floquet reflection coefficients $R_\sigma(\omega_\text{d},\omega_\text{d};K_\parallel)$. In the main text, we particularize the above formula for $d_x,\text{ }d_y = 0$, which maximizes the coupling to TM-modes in the near-field.
\subsection{Near-Field Gain: Dipole Radiation in the Quasistatic Regime}
In the main text, we consider $\Omega = 2\omega_\text{ENZ}$ and $\alpha = 0.4$. For the case in which $\omega_\text{d} = \Omega - \omega_\text{SP}$, with the dipole being resonant with the surface plasmon replica, we have increasing gain, in the form of a negative value of the total power, as the distance between the emitter and the surface decreases. We further argue that the gain contribution to the total power, $P_\text{gain}$, becomes truly large at the onset of the \textit{quasistatic} regime. In static media, the quasistatic regime refers to the $\omega_\text{d}/c_0\ll 1 / h$ scenario, in which the main contribution to the total power does not come from neither the radiative modes nor the poles of the structure, but from non-radiative losses. This is often referred to as the quenching of the spontaneous emission \cite{novotny2012principles}. Crucially, while the poles contribution to the total power follows an exponential law $\exp(-2k_\text{pole}h)$, the quasistatic one goes with $h^{-3}$.
\\
\begin{figure}[H]
\makebox[\columnwidth][c]{
    \centering
    \hspace{0 mm}
    \includegraphics[scale = 0.55]{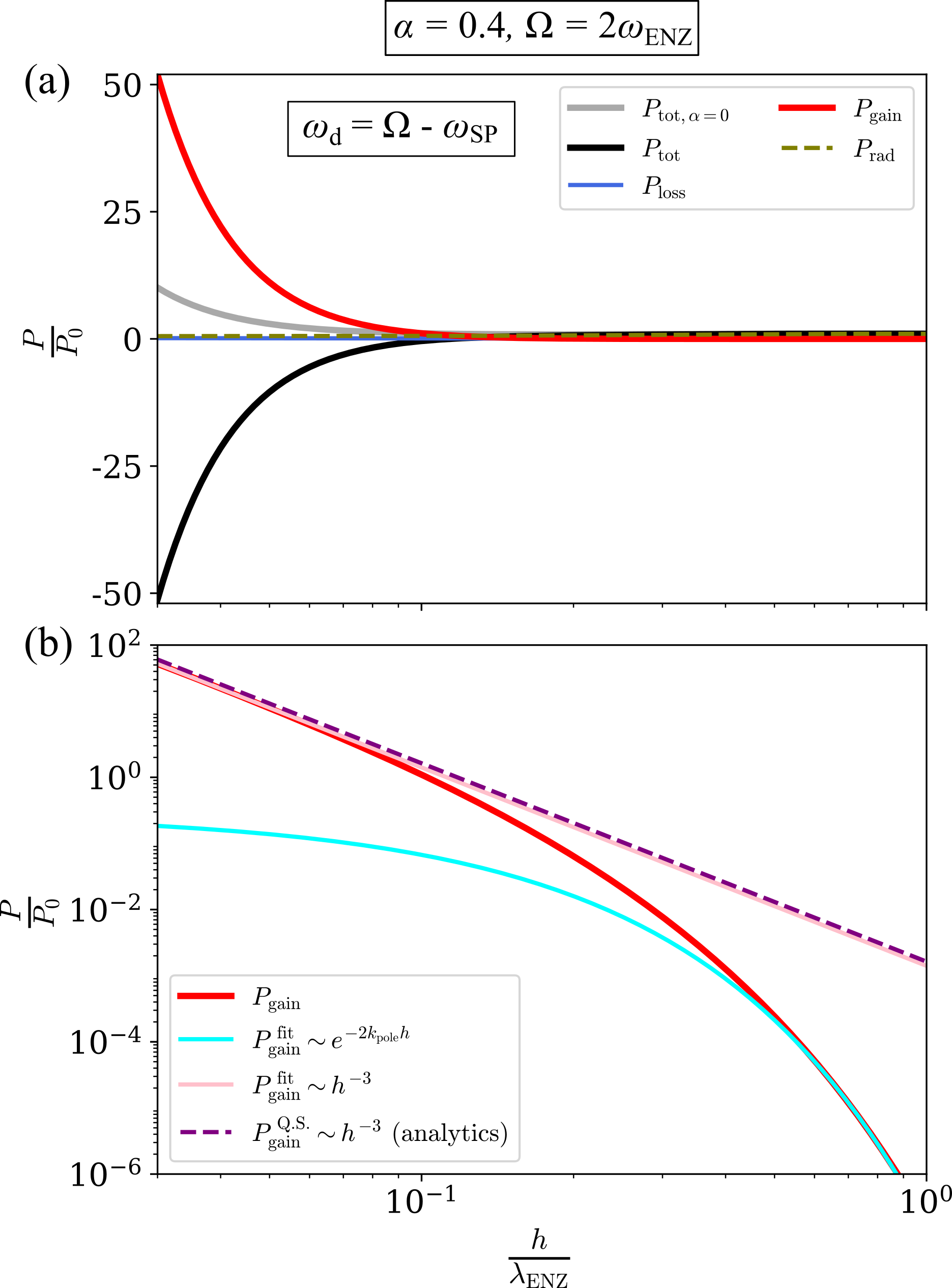} }
    \caption{(a) Total power for the static (solid black line) and time-modulated cases (solid orange), the latter for $\Omega = 2\omega_\text{ENZ}$ and $\alpha = 0.4$, and for $\omega_\text{d} = \Omega - \omega_\text{SP}$. We also plot the different contributions to the total power for the sake of completeness, as done in the main text: $P_\text{loss}$ in solid blue, $P_\text{gain}$ in solid red and $P_\text{rad}$ in dashed green. In (b), we fit $P_\text{gain}$ to an exponential (solid cyan) and an algebraic $h^{-3}$ curve (solid pink); we also plot the approximate analytical formula for the quasistatic contribution (dashed purple), which matches well to the numerics.}
    \label{fig:SM_fig1} 
\end{figure}
We now make two assumptions: (i) $\alpha\ll1$, which allows us to work with a minimal model, in which only two bands are taken into account (the fundamental one $n = 0$ and its first negative replica $n=-1$), and (ii) $\omega_\text{d}/c_0\ll1/h$, the quasistatic approximation. The quasistatic approximation holds only for small enough $h$, and in principle, enables us to treat the metallic slab as a metallic half-space. This in turn greatly simplifies the scattering matrix analytical calculations and allows us to find an approximate formula for the frequency conserving Fresnel-Floquet reflection coefficient. The latter is then given by:

\begin{equation}
    R_{00} = R^{(0)}_0 + \left(\frac{\alpha}{2}\right)^2R^{(2)}_{00},
\end{equation}
where $R^{(0)}_0 = R^{(0)}_\text{TM}(\omega,K_\parallel)$ and $R^{(2)}_{00}$ is the second order correction (and the lowest-order one) to the frequency conserving reflection coefficient, which reads
\begin{equation}
    \begin{split}
        R^{(2)}_{00} & = \Bigg[-\frac{1}{2}\left(1 - (R^{(0)}_0)^2\right)\frac{\chi^{(0)}_0}{\varepsilon^{(0)}_0}\frac{\chi^{(0)}_{-1}\left(\frac{\omega_{-1}}{c_0}\right)^2}{\Delta^{(0)}} + \\& \frac{1}{4}\left(1 + R^{(0)}_0\right)^2\left(1 + R^{(0)}_{-1}\right)\chi^{(0)}_0\Big(1-(1 + \varepsilon_0^{(0)})\frac{\left(\frac{\omega_0}{c_0}\right)^2}{\Delta^{(0)}}\Big)\chi^{(0)}_{-1}\Big(1+(1 + \varepsilon_{-1}^{(0)})\frac{\left(\frac{\omega_{-1}}{c_0}\right)^2}{\Delta^{(0)}}\Big) + \\&
        \frac{1}{2}(1 + R^{(0)}_0)(1 + R^{(0)}_{-1})\frac{\chi^{(0)}_0\left(\frac{\omega_0}{c_0}\right)^2}{\Delta^{(0)}}\chi^{(0)}_{-1}\Big(1+(1 + \varepsilon_{-1}^{(0)})\frac{\left(\frac{\omega_{-1}}{c_0}\right)^2}{\Delta^{(0)}}\Big)\Bigg].
    \end{split}
    \label{eq:refQS}
\end{equation}

In the above, we have used the following short-hand notations: $\chi^{(0)}_0 = \chi^{(0)}(\omega)$, $\varepsilon^{(0)}_0 = \varepsilon_\infty +  \chi^{(0)}(\omega)$, $\chi^{(0)}_{-1} = \chi^{(0)}(\omega - \Omega)$, $\varepsilon^{(0)}_{-1} = \varepsilon_\infty +  \chi^{(0)}(\omega - \Omega)$, $R^{(0)}_{-1} = R^{(0)}_\text{TM}(\omega - \Omega,K_\parallel)$, and lastly, $\Delta^{(0)} = \varepsilon^{(0)}_0\left(\frac{\omega_0}{c_0}\right)^2 - \varepsilon^{(0)}_{-1}\left(\frac{\omega_{-1}}{c_0}\right)^2$. Crucially, in the definitions of $\chi^{(0)}_0$ and $\chi^{(0)}_{-1}$ we have not included the contribution of the background permittivity, $\varepsilon_\infty$, to keep the notation simple in Eq.~\eqref{eq:refQS}. In the $K_\parallel\gg\omega_\text{d}/c_0$ limit, Eq.~\eqref{eq:refQS} becomes independent of $K_\parallel$ and thus, we can analytically integrate the total power in the quasistatic regime, which is given by:
\begin{equation}
    \frac{P_\text{tot}}{P_0} = 1 + \frac{3c_0^3}{8\omega_\text{d}^3}\frac{\mathfrak{I}\{R_{00}\}}{h^3}.
    \label{eq:pradQS}
\end{equation}
In Fig.~\ref{fig:SM_fig1}(a), we plot the total power for the static and time-modulated cases, for values of $\Omega = 2\omega_\text{ENZ}$ and $\alpha = 0.4$. We see $P_\text{gain}$ (red solid line, as seen in the legend) continues to increase as $h$ decreases, reaching values of $P_\text{gain}>50P_0$ for $h = 3\cdot10^{-2}\lambda_\text{ENZ}$. Correspondingly, we have $P_\text{tot}<-50P_0$ (black solid line) and the quenching of the radiated power is transformed into non-radiative gain, which dominates the physics of the near-field.
\\
In Fig.~\ref{fig:SM_fig1}(b), we plot $P_\text{gain}$ and fit its values along appropriate intervals of $h$ to an exponential (cyan solid line) and an algebraic curve $h^{-3}$ (solid pink line). We also plot the absolute value\footnote{We take the absolute value because the quasistatic contribution to the total power is negative for $\omega = \Omega - \omega_\text{SP}$, and because we wish to compare it to $P_\text{gain}$, which is defined to be positive.} of the second term in Eq.~\eqref{eq:pradQS}, the  approximate analytical formula (for $\alpha\ll1$ and $\omega_\text{d}/c_0\ll1/h$) for the quasistatic regime contribution to the total power (purple dashed line). We see $P_\text{gain}$ fits perfectly well to the exponential down to $h\sim5\cdot10^{-1}\lambda_\text{ENZ}$, at which $P_\text{gain}<10^{-2}P_0$. Hence, the replica of the surface plasmon polariton does not yield a large value of $P_\text{gain}$ and thus, the negative values of the total power (for small enough $h$) do not come from the pole. For $h<5\cdot10^{-1}\lambda_\text{ENZ}$, the fitting to the algebraic curve $h^{-3}$ works much better, becoming perfect from $h \lesssim 10^{-1}\lambda_\text{ENZ}$. Therefore, it is indeed the replica of the surface plasmon that is responsible for the large values of $P_\text{gain}$ and thus, of the dipole absorbing electromagnetic energy in the near-field. We can also see that the numerics (solid pink) and the analytics (dashed purple) match each other almost completely, despite the fact that the analytical formula was derived assuming $\alpha\ll1$.
\\
Lastly, we briefly comment on non-local effects in the linear response of the slab, which we have neglected. For ITO, the parameter $\beta$, which determines the strength of non-local effects, is $\beta = 0.0057c_0$ \cite{thouin2025field}. Non-local effects start to become important when $K_\parallel \sim \omega / \beta = K^\text{(non-loc)}_\parallel$. From the latter, we can define the (frequency-dependent) height $h^\text{(non-loc)} = 1 / K^\text{(non-loc)}_\parallel$ at which the emitter begins to see non-local effects. For $\omega_\text{d} = \Omega - \omega_\text{SP}$ and $\Omega = 2\omega_\text{ENZ}$,  we have $h^\text{(non-loc)} \sim 10^{-3}\lambda_\text{ENZ}$. Thus, for smaller dipole-slab distances than considered in this work, one would need to take into account non-local effects in the plasma response.

\subsection{Far-Field Control: Leaky resonance dependence with $\alpha$ and $\Omega$ }
In the main text, we show that the temporal modulation caused the Brewster condition and its negative frequency replica to interact at the parametric resonance condition, $\omega = 0.5\Omega = \omega_\text{ENZ}$. As a result of this interaction, a sharp peak is formed inside the lightcone, at $\omega = \omega_\text{ENZ}$. In this section, we study the dependence of this peak with the modulation strength $\alpha$, and with the modulation frequency $\Omega$.
\begin{figure}[H]
\makebox[\columnwidth][c]{
    \centering
    \hspace{0 mm}
    \includegraphics[scale = 0.45]{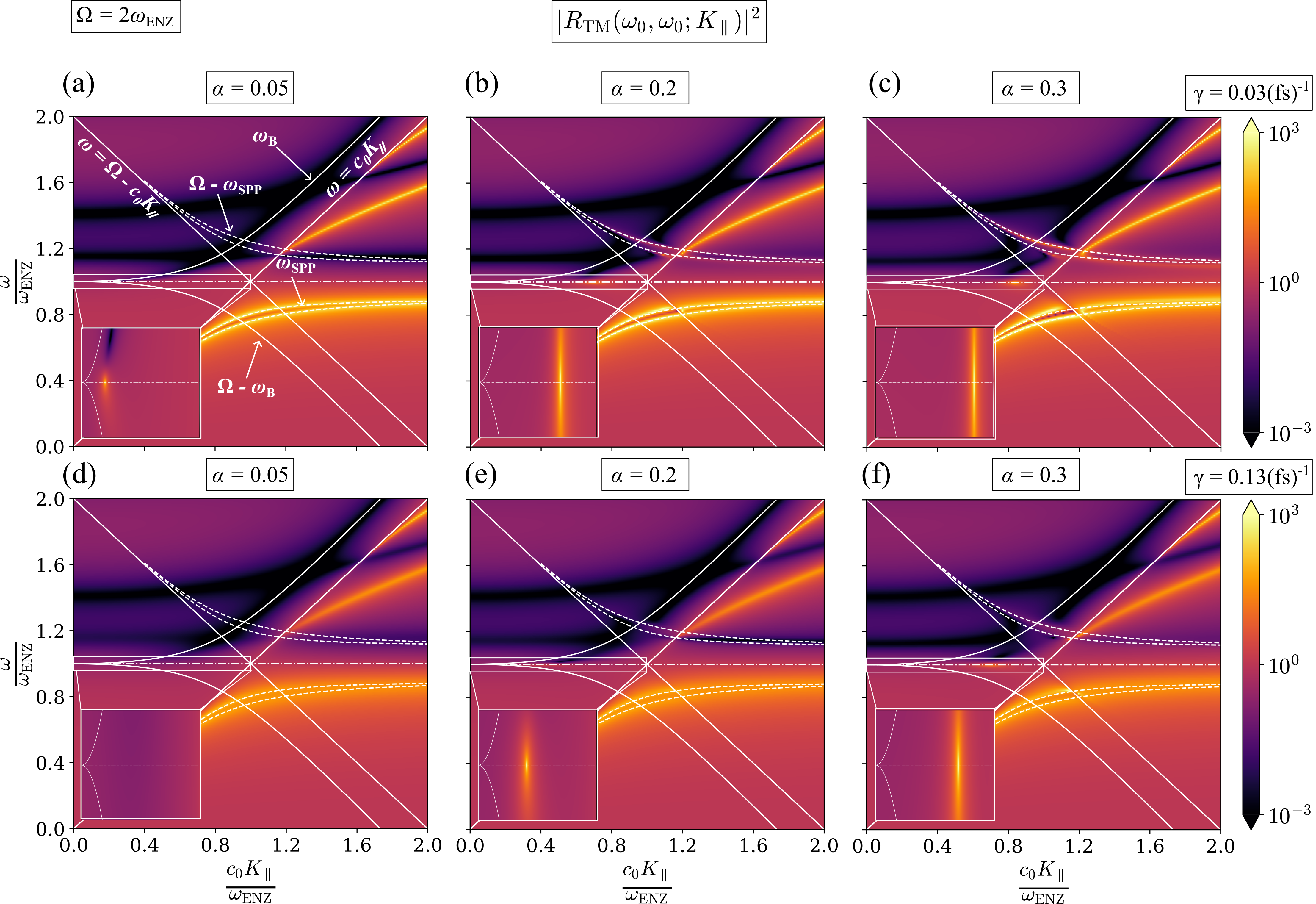} }
    \caption{Time-modulated reflectance for TM-polarized waves, for fixed $\Omega = 2\omega_\text{ENZ}$ and for varying modulation strength: $\alpha=0.05$ (a)-(d), $\alpha=0.2$ (b)-(e), $\alpha=0.3$ (c)-(f). Insets centered around $\omega = 0.5\Omega = \omega_\text{ENZ}$ show the sharp peak inside the lightcone progressively moving towards larger values of $K_\parallel$ as $\alpha$ increases. For (a)-(c), we work with $\gamma = 0.03\text{(fs)}^{-1}$, while for (d)-(f) it is $\gamma = 0.13\text{(fs)}^{-1}$, as in the main text.} 
    \label{fig:SM_fig2} 
\end{figure}
\par
In Fig.~\ref{fig:SM_fig2}, we plot the time-modulated reflectance for TM polarization, for a fixed value of $\Omega = 2\omega_\text{ENZ}$, and for different values of $\alpha$. For Figs.~\ref{fig:SM_fig2}(a)-(c) we set the damping rate to $\gamma = 0.03\text{(fs)}^{-1}$, whereas for Figs.~\ref{fig:SM_fig2}(d)-(f), we take the same value as in the main text, $\gamma = 0.13\text{(fs)}^{-1}$. Therefore, in the first row of Fig.~\ref{fig:SM_fig2}, losses are approximately a 25$\%$ of those used in the main text and in the second row.
\\
In Fig.~\ref{fig:SM_fig2}(a) we take $\alpha = 0.05$. The inset, centered around $\omega = 0.5\Omega = \omega_\text{ENZ}$, shows the sharp leaky resonance inside the lightcone, to whose left lies the crossing point between $\omega_\text{B}$ and $\Omega - \omega_\text{B}$. Still looking at the inset, on top of the peak we can see a dark purple line. This actually corresponds to the Brewster condition, which deviates from its static dispersion relation, given by $\omega_\text{B}$, as a result of the time modulation.  Increasing $\alpha$ pushes the peak to larger values of $K_\parallel$, as seen in Figs.~\ref{fig:SM_fig2}(b)-(c). This resembles the opening of a $K_\parallel-$gap, with the peak inside the lightcone behaving as the edge of the gap: as the modulation strength increases, the gap becomes wider and the edge moves to larger values of the lateral momentum. Additionally, as $\alpha$ increases, the leaky resonance becomes brighter and broader in frequency.
\\
We now focus on the second row of Fig.~\ref{fig:SM_fig2}, in which losses are higher than in the first one. A glance at the inset of Fig.~\ref{fig:SM_fig2}(d) shows that the leaky resonance is absent for $\alpha = 0.05$. This is in sharp contrast with the case of Fig.~\ref{fig:SM_fig2}(a), in which for the same value of $\alpha$, the leaky resonance is indeed present. Increasing $\alpha$ eventually makes the leaky resonance appear, as seen in Figs.~\ref{fig:SM_fig2}(e)-(f). However, it appears at a lower value of $K_\parallel$ and is less broad in frequency than for the low-loss case [see Figs.~\ref{fig:SM_fig2}(b)-(c)]. Therefore, losses are seen to determine the strength of the interaction between $\omega_\text{B}$ and $\Omega - \omega_\text{B}$, and thus set the threshold for the appearance of the leaky resonance at $\omega\sim0.5\Omega = \omega_\text{ENZ}$.
 
\begin{figure}[H]
\makebox[\columnwidth][c]{
    \centering
    \hspace{0 mm}
    \includegraphics[scale = 0.45]{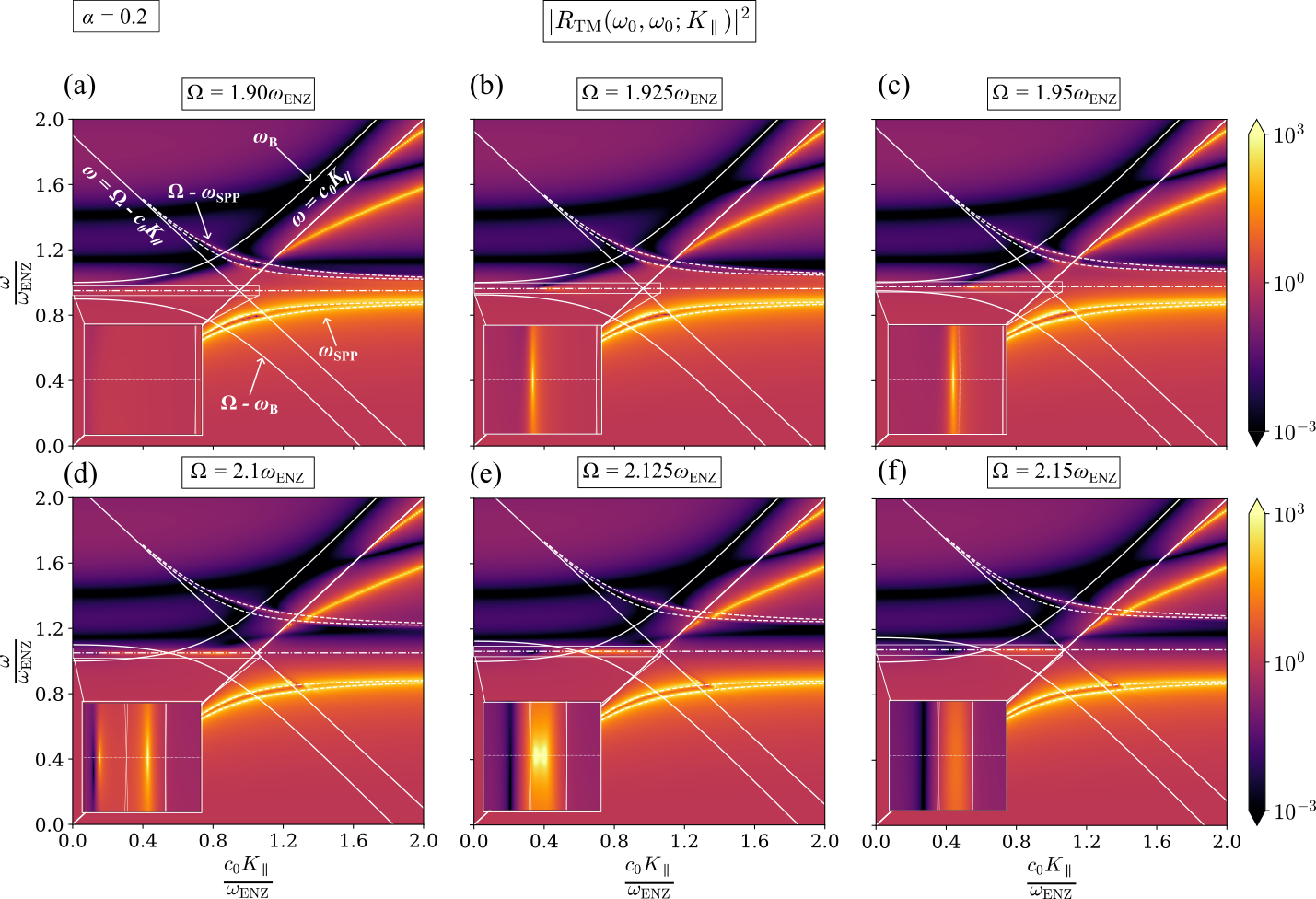} }
    \caption{Time-modulated reflectance for TM-polarized waves, for fixed $\alpha = 0.2$ and for varying modulation frequency: $\Omega=1.9\omega_\text{ENZ}$ (a), $\Omega=1.925\omega_\text{ENZ}$ (b), $\Omega=1.95\omega_\text{ENZ}$ (c), $\Omega=2.1\omega_\text{ENZ}$ (d), $\Omega=2.125\omega_\text{ENZ}$ (e) and $\Omega=2.15\omega_\text{ENZ}$ (f). Insets are centered around $\omega = 0.5\Omega$.}
    \label{fig:SM_fig3} 
\end{figure}
\par
In Fig.~\ref{fig:SM_fig3} we keep the modulation strength fixed to $\alpha = 0.2$, and consider different values of the modulation frequency. We also work with $\gamma = 0.03\text{(fs)}^{-1}$ to better visualize the results. The insets are centered around the parametric resonance condition, $\omega = 0.5\Omega$. A glance at Fig.~\ref{fig:SM_fig3}(a) shows that for $\Omega = 1.90\omega_\text{ENZ}$, there is no resonance inside the lightcone. As $\Omega$ increases, the resonance is formed and moves towards larger values of $K_\parallel$ [see the insets in Fig.~\ref{fig:SM_fig3}(b)-(c)]. For $\Omega = 2.1\omega_\text{ENZ}$, a second peak appears to the left of the original one. Further increasing the modulation frequency shows that the two peaks approach each other, until they eventually coalesce and disappear [see Fig.~\ref{fig:SM_fig3}(f)]. Therefore, we can see that for large enough detuning, whether positive ($\Omega >2\omega_\text{ENZ}$) or negative ($\Omega <2\omega_\text{ENZ}$), the peak or peaks present inside the lightcone end up disappearing. The appearance of a second sharp resonant mode for large enough $\Omega$ resembles the closing of a $K_\parallel-$gap: as the detuning (with respect to $\Omega = 0.5\omega_\text{ENZ}$) increases, the interaction between $\omega_\text{B}$ and $\Omega - \omega_\text{B}$ weakens. Correspondingly, the two edges of the $K_\parallel-$gap approach each other, coalescing as the gap closes and eventually disappearing. 

\bibliographystyle{unsrt}
\bibliography{SupMat}
%\printbibliography %Prints bibliography